\documentclass[preprint,12pt,3p]{elsarticle}
\usepackage[utf8]{inputenc} 
\usepackage[T1]{fontenc}    
\usepackage{hyperref}       
\usepackage{url}            
\usepackage{booktabs}       
\usepackage{amsfonts}       
\usepackage{nicefrac}       
\usepackage{microtype}      
\usepackage{lipsum}
\usepackage{amsmath,amssymb,graphicx}
\usepackage{float}
\usepackage{siunitx}
\usepackage{array}
\usepackage{lmodern}
\usepackage{amssymb}
\usepackage{xcolor}
\usepackage{multicol}
\usepackage{multirow}
\usepackage{orcidlink}
\usepackage{amsmath,amssymb,exscale}
\usepackage{amsmath}
\usepackage{blindtext}
\usepackage{hyperref}
\journal{Elsevier}
\begin{document}
\begin{frontmatter}
\title{Elasticity of Irida-graphene-based and Sun-graphene-based nanotubes: A study by fully atomistic reactive classical molecular dynamics simulations}

\author[Damghan University]{ Reza Kalami\corref{author}\,\orcidlink{0000-0003-1194-4210}
}
\cortext[author]{Corresponding authors:\\ rezakk01@gmail.com (R. Kalami), ramon@fisica.ufmt.br (R. S. Ferreira), saketabi@du.ac.ir (S. A. Ketabi), josemoreiradesousa@ifpi.edu.br(J. M. De Sousa).}

\author[UFMT]{Ramon S. Ferreira\corref{author}\,\orcidlink{0000-0002-3421-6278}
}

\author[Damghan University]{Seyed  Ahmad Ketabi\corref{author}\,\orcidlink{0000-0001-6629-0918}
}

\author[IFPI]{José M. De Sousa \corref{author}\,\orcidlink{0000-0002-3941-2382}
}

\affiliation[Damghan University]{organization={School of Physics},
            addressline={Damghan University}, 
            city={Damghan},
        country={Iran}}
     
\affiliation[UFMT]{organization={Instituto de Física},
            addressline={Universidade Federal de Mato Grosso}, 
            city={Cuiabá},
            postcode={78060-900}, 
            state={Mato Grosso},
            country={Brazil}}  


\affiliation[IFPI]{organization={Instituto Federal de Educa\c c\~ao, Ci\^encia e Tecnologia do Piau\'i -- IFPI},
            addressline={Primavera}, 
            city={São Raimundo Nonato},
            postcode={64770-000}, 
            state={Piauí},
            country={Brazil}} 
                 
\begin{abstract}
In this work, we present a theoretical investigation of the elasticity properties of new exotic carbon allotropes studied recently called Irida-graphene and Sun-graphene. The Irida-graphene is a new 2D all-$sp^{2}$ carbon allotrope composed of fused rings containing 3-6-8 carbon atoms. However, The Sun-graphene is a new carbon allotrope analogous to graphene, has a semi-metal nanostructure and presents two Dirac cones in its band structure. Despite being theoretically studied in single-layer ($2D$), no study of the elastic properties of Irida-graphene-based nanotubes (Irida-CNTs) and Sun-graphene-based nanotubes (Sun-CNTs) has not been studied yet. Thus, we seek to investigate the elastic properties of nanotubes of these new exotic allotropes of carbon in quasi-one-dimensional geometry ($1D$), nanotubes for different chiralities, diameters and lengths at room temperature. For the development of the theoretical investigation of the elasticity of the Irida-CNTs and Sun-CNTs, we developed a fully reactive (ReaxFF) atomistic classical molecular dynamics simulation method. Our theoretical results obtained in the computational simulations show that the Irida-CNTs nanotubes behave as flexible nanostructure and have a slightly higher value of strain rate than Sun-CNTs. The
values of Young’s modulus for Irida-CNTs are slightly larger range ($640.34$ GPa - $825.00$
GPa), while the Young’s Modulus range of Sun-CNTs is $200.44$ GPa - $472.84$ GPa, the lower values
than those presented for conventional carbon nanotubes (CNT). These findings provide insights into the mechanical behavior of Irida-CNTs and Sun-CNTs and their potential applications in nanoscale devices.
\end{abstract}

\begin{keyword}
Classical molecular dynamics method \sep ReaxFF \sep elastic properties \sep nanofracture pattern \sep Irida-graphene \sep  Sun-graphene \sep nanotubes 
\end{keyword}
\end{frontmatter}

\section{Introduction}
\label{INT}

Using one-dimensional nanostructures in theoretical and experimental studies has received a lot of attention from the world scientific community, since the synthesis of the one-dimensional tubular morphology known today as carbon nanotube (CNT) \cite{iijima1991helical,jin2022cagdal3o7,campiglio2011quasi,herrera2021theoretical}. Theoretically and experimentally investigations of CNTs are a new subject for all materials \cite{dresselhaus1998physical,reich2008carbon,ebbesen1994carbon}. After the synthesis of quasi- $1D$ carbon nanotubes, there was a renewed interest in the development of basic theoretical and experimental research, mainly increasing the technological potential in the development of traditional carbon fibers \cite{salvetat2002mechanical,bhattacharyya2008carbon}. 
Thus, significant scientific challenges from a technological point of view, the large-scale synthesis of high-purity carbon nanotube samples and their manipulation at the nanometric scale are not yet fully understood. Similarly to $2D$, quasi-$1D$ carbon nanotubes have received particular attention in the
last decades. Thus, understanding and developing theoretical and experimental findings is fundamental for the understanding of physical and chemical properties of carbon nanotubes for different applicability and sustained development of new structures and new materials \cite{zhang2010functional,navrotskaya2020hybrid,legoas2003molecular}. CNTs are suitable for application in biomedical devices \cite{coelho2015carbon,sharma2016biomedical,saliev2019advances},  as field-emission electron sources \cite{de1995carbon}, tissue scaffolds \cite{harrison2007carbon}, actuators \cite{baughman1999carbon} and, artificial muscles \cite{lima2012electrically}, among other applications that can be researched in the literature. A single-walled carbon nanotube (SWCNT) can be thought of as formed by rolling a sheet of graphene to make a seamless cylinder, where the way it is wound determines its chirality like armchair $(n,n)$, zigzag $(n, 0)$ and chiral $(n,m)$. Thus, CNTs can be considered as single-walled and multi-walled cylindrical shapes (MWCNT) \cite{dresselhaus1998physical}. The SWNT should be metallic when $(n - m)/3$  is an integer and otherwise, they are semiconductors \cite{ouyang2002fundamental}. The Young's modulus of SWCNT is on the order of $1000$ GPa \cite{lu1997elastic}. 

Due to the versatile nature of carbon, new exotic forms of CNTs are also studied, where their results will serve as a theoretical library for future experimental applications. Graphyne, introduced in $1987$ by Baughman, Eckhardt, and Kertesz, is a two-dimensional carbon allotrope composed of carbon atoms in both $sp$ and $sp^{2}$ hybridized states that link benzenoid rings.\cite{baughman1987structure}. New families of carbon single-walled nanotubes called graphyne-based nanotubes (GNTs) were studied recently using tight-binding and \textit{ab initio} density functional methods. Proposed by Coluci, V. R., \textit{et al.} (2003), GNTs is metallic for armchair and has metallic or semiconducting behavior for zigzag one \cite{coluci2003families,coluci2004theoretical}. De Sousa, J. M. \textit{et al.} (2019) studied the elasticity of GNTs under uniaxial
tensile stress performed by fully atomistic reactive (ReaxFF) classical molecular dynamics simulations and density functional theory (DFT) calculations methods, respectively. The results showed that GNTs nanofracture occurs at larger strain values in comparison to conventional CNTs, but with smaller ultimate tensile strength (UTS), $44 - 49$ GPa and Young’s modulus values range of the $149 - 472$ GPa \cite{de2019elastic}. De Sousa, J. M. \textit{et al.} (2016) studied the mechanical behavior of GNTs under torsional strains performed by fully classical reactive (ReaxFF) molecular dynamics simulations method. The results showed that GNTs are more flexible than conventional CNTs and with nanofracture angles that are up to 35 times higher than CNTs \cite{de2016torsional}. Penta-graphene, another new carbon allotrope proposed by Zhang, S. et al. (2015), is a type of carbon allotrope that exists in a 2D form. It is made up of pentagons of carbon atoms and has a structure that resembles the Cairo pentagonal tiling pattern \cite{zhang2015penta}.  De Sousa, J. M. \textit{et al.} (2021), conducted a research on the elastic properties of nanotubes created by folding penta-graphene membranes, which are known as penta-graphene nanotubes (PGNTs) by reactive (ReaxFF) classical molecular dynamics simulations and DFT methods, respectively. The elasticity parameters of PGNTs such as Young’s modulus values of $680-800$ GPa, the critical strain of $18-21$\% and the ultimate tensile strength of $85-110$ GPa were obtained  \cite{de2021mechanical}. Proposed by Wang, S.\textit{ et al.} (2018), the POPgraphene is a recently discovered type of carbon allotrope that consists of rings arranged in a $5-8-5$ structure\cite{wang2018popgraphene}. Brandão, W. H. S. \textit{et al.}
 (2021), used a method called reactive (ReaxFF) classical molecular dynamics simulations to examine the mechanical properties of nanotubes made from POPgraphene. It's Young's modulus is a range of the ($750-900$ GPa) and the ultimate strength is ($120-150$ GPa) \cite{brandao2021mechanical}. In addition, Wang, Z. et al. (2015) proposed a new graphene allotrope named Phagraphene,  penta-hexa-hepta-graphene, which is a new carbon nanostructure composed of $5-6-7$ carbon rings \cite{wang2015phagraphene}. Júnior, M. L. P. \textit{et al.} (2020), investigated the elasticity properties of phagraphene nanotubes (PhaNTs) by fully atomistic reactive (ReaxFF) classical molecular dynamics (MD) simulations method. The results for the Young's Modulus obtained $706.15 - 796.75$ GPa, the ultimate tensile strength is $169.58 - 170.33$ GPa and the critical strain range is $17.76 - 22.24$\%~ \cite{junior2020elastic}. 
 
 Given that the experimental results show that graphene \cite{novoselov2004electric} is a semiconductor with zero energy gap \cite{withers2010electron}, the new allotropes have been investigated theoretically with the objective of obtaining new nanostructures with qualities equal to or superior to graphene. For example, new allotropes of carbon have been theoretically proposed recently such as Irida-graphene and Sun-graphene (2023). Irida-graphene is a new 2D all-$sp^{2}$ carbon allotrope composed of fused rings containing $3-6-8$ carbon atoms. Proposed by Júnior, M. L. P. \textit{et al.} (2023) and performed by reactive (ReaxFF) classical molecular dynamics simulations and density functional theory (DFT) the findings showed that Irida-graphene is a nanostructure that is both metallic and non-magnetic. It has a Dirac cone located at the center of its band structure, just above the Fermi level. In addition, it has an elastic modulus of approximately $396$ GPa and also photon energies of about $3.0$ eV \cite{junior2023irida}. On the other hand, Sun-graphene is a new $2D$ carbon allotrope ($8-16-4$ Graphyne) which was proposed by Tromer, R. M. \textit{et al.} (2023). It is a semi-metal nanostructure and presents two Dirac cones in its band structure with a formation energy of about $-8,57$ eV/atom and Young's Modulus of the $262.37$ GPa which obtained by reactive (ReaxFF) classical molecular dynamics simulations and DFT \cite{tromer2023sun}. Despite being theoretically proposed in two-dimensional geometry recently, their cylindrical morphology in nanotubes has not been studied yet.

Therefore, using the state of art theoretical tools such as computer simulation by reactive (ReaxFF) classical molecular dynamics simulations to explore the nanostructural, dynamics elasticity properties of Irida-graphene-based (Irida-G-NTs) and Sun-graphene-based (Sun-G-NTs) nanotubes is the aim of this work. Here, we performed computer simulations to estimate the mechanical properties and understand the fracture patterns of Irida-G-NTs and Sun-G-NTs with zigzag $(n,0)$, armchair $(n,n)$ chiralities and chiral $(n,m)$. Our results show that Irida-CNTs nanotubes behave as flexible nanostructures and have a slightly higher value of strain rate than Sun-CNTs due to nanostructural reconstructions. The
values of Young’s modulus for Irida-CNTs are slightly larger range ($640.34$ GPa - $825.00$
GPa) and Sun-CNTs, the Young’s Modulus range ($200.44$ GPa - $472.84$ GPa). 
In this section, we briefly introduced the preliminary concepts and the motivations for the development of this theoretical work. The computational methodology used to
model the nanostructures of Irida-CNTs and Sun-CNTs are described in Sections 2 and 3, the results and discussion are presented in Section 4 followed by conclusions
and remarks in Section 5

\section{Generation of Irida-graphene-based and Sun-graphene-based nanotubes (Irida-CNTs and Sun-CNTs)}

\begin{figure}[hp!]
\begin{center}
\includegraphics[width=\columnwidth]{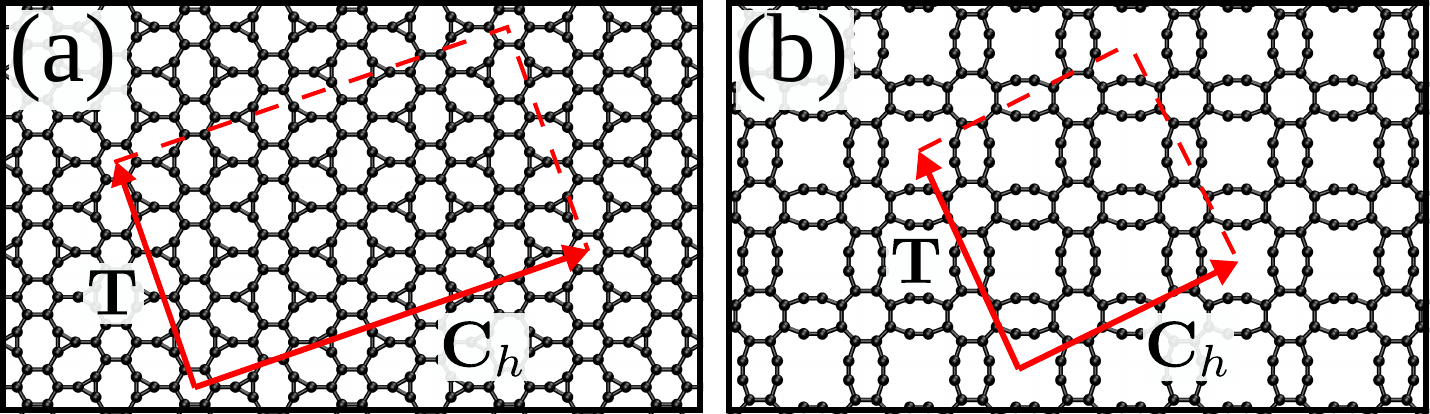}
\caption{\footnotesize{(a) Illustration of the Irida-graphene structure and a representation of the chiral ($\mathbf{C}_h$) and translational ($\mathbf{T}$) vectors for a $(4,1)$ Irida-CNT. (b) Representation of the Sun-graphene layer and the corresponding chiral and translational vectors for a $(2,1)$ Sun-CNT.}}
\label{fig-tubes}
\end{center}
\end{figure}

The details involved in the determination of the atomic structure of Irida-CNTs and Sun-CNTs differ significantly from each other. This is not only because they are based on different 2D structures, but also because these parent structures have different symmetries. On one hand, Irida-graphene is organized in a hexagonal 2D lattice, with symmetries described by the wallpaper group, similar to graphene. In this way, the definition of its structure regarding chiral, translational vectors and chirality are completely analogous to that of graphene. Namely, the chiral vector ($\mathbf{C}_h$) is defined in terms of the lattice vectors of Irida-graphene ($\mathbf{a}_1$ and $\mathbf{a}_2$) by
\begin{equation}
\mathbf{C}_h=n\mathbf{a}_1+m\mathbf{a}_2
\end{equation}
where $n$ and $m$ are integers with $n>0$ and $0\le m\le$. The condition on $m$ ensures we consider all the distinct tube geometries allowed by the hexagonal symmetry. In other words, $0\le m\le$ ensures the nanotube's chiralities lie in the $0^\circ$-to-$30^\circ$ range for the tube chirality, which is the angle between $\mathbf{a}_1$ and $\mathbf{C}_h$, given by
\begin{equation}
\theta=\cos^{-1}\frac{2n+m}{2\sqrt{n^2+nm+m^2}}.
\end{equation}
As well known, the translational vector ($\mathbf{T}$) is given by
\begin{equation}
\mathbf{T}=\frac{2m+n}{d}\mathbf{a}_1-\frac{2n+m}{d}\mathbf{a}_2,
\end{equation}
with $d$ being the greatest common divisor between $2m+n$ and $2n+m$. As an example, we show these vectors for the case of a $(4,1)$ tube on Fig.~\ref{fig-tubes}a. We note the $(n,n)$ and $(n,0)$ Irida-G-NTs do not feature an exact armchair or zigzag line of atoms along their cross-sections. However, we will maintain the same convention used for graphene nanotubes and call them armchairs and zigzags Irida-CNTs, respectively.

On the other hand, Sun-graphene is set over a tetragonal (or square lattice). Since the angle between $\mathbf{a}_1$ and $\mathbf{a}_2$ is $90^\circ$, the nanotube chirality is now given 
\begin{equation}
\theta=\cos^{-1}\frac{n}{\sqrt{n^2+m^2}}
\end{equation}
and it now ranges from $0^\circ$ to $45^\circ$, so that we keep the $0\le m\le$ condition. Its translational vector is now given by
\begin{equation}
\mathbf{T}=\frac{m}{d}\mathbf{a}_1-\frac{n}{d}\mathbf{a}_2,
\end{equation}
with $d$ being the greatest common divisor between $n$ and $m$. Just as a matter of convention, we will also coin the $(n,n)$ and $(n,0)$ Sun-CNTs as armchair and zigzag nanotubes, respectively. In Fig.~\ref{fig-tubes}b we illustrate some structural details for the $(2,1)$ Sun-CNT.

\begin{figure}[ht!]
\begin{center}
\includegraphics[angle=0,scale=0.28]{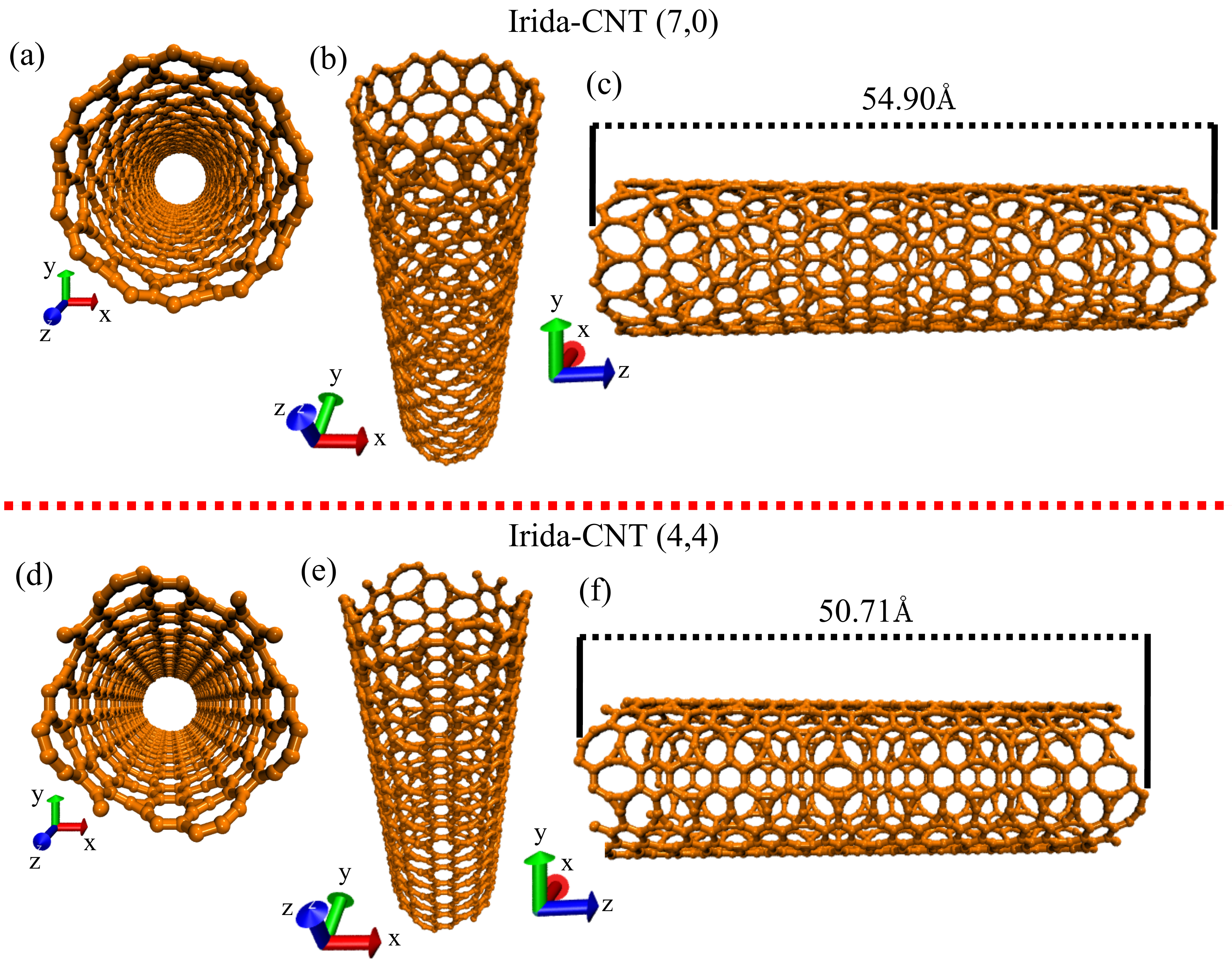}
\caption{\footnotesize{Representative nanostructural fully atomic model of Irida-graphene-based nanotube (Irida-CNT) zigzag $(7,0)$ and armchair $(4,4)$. (a) Front view in the van der Waals and dynamics bonds representation of Irida-CNT $(7,0)$, (b) longitudinal view of the nanotube and (c) the side view of the nanotube. (d) Front view in the van der Waals and dynamics bonds representation of Irida-CNT $(4,4)$, (e) longitudinal view of nanotube and (f) the side view of the nanotube. Information on the radius, number of atoms, diameter and length of all Irida-CNTs studied in this research work can be viewed in the  table \ref{tab:Irida-CNT;Sun-CNT}.}}
\label{FIG:Irida-CNT:01}
\end{center}
\end{figure}

\begin{figure}[ht!]
\begin{center}
\includegraphics[angle=0,scale=0.28]{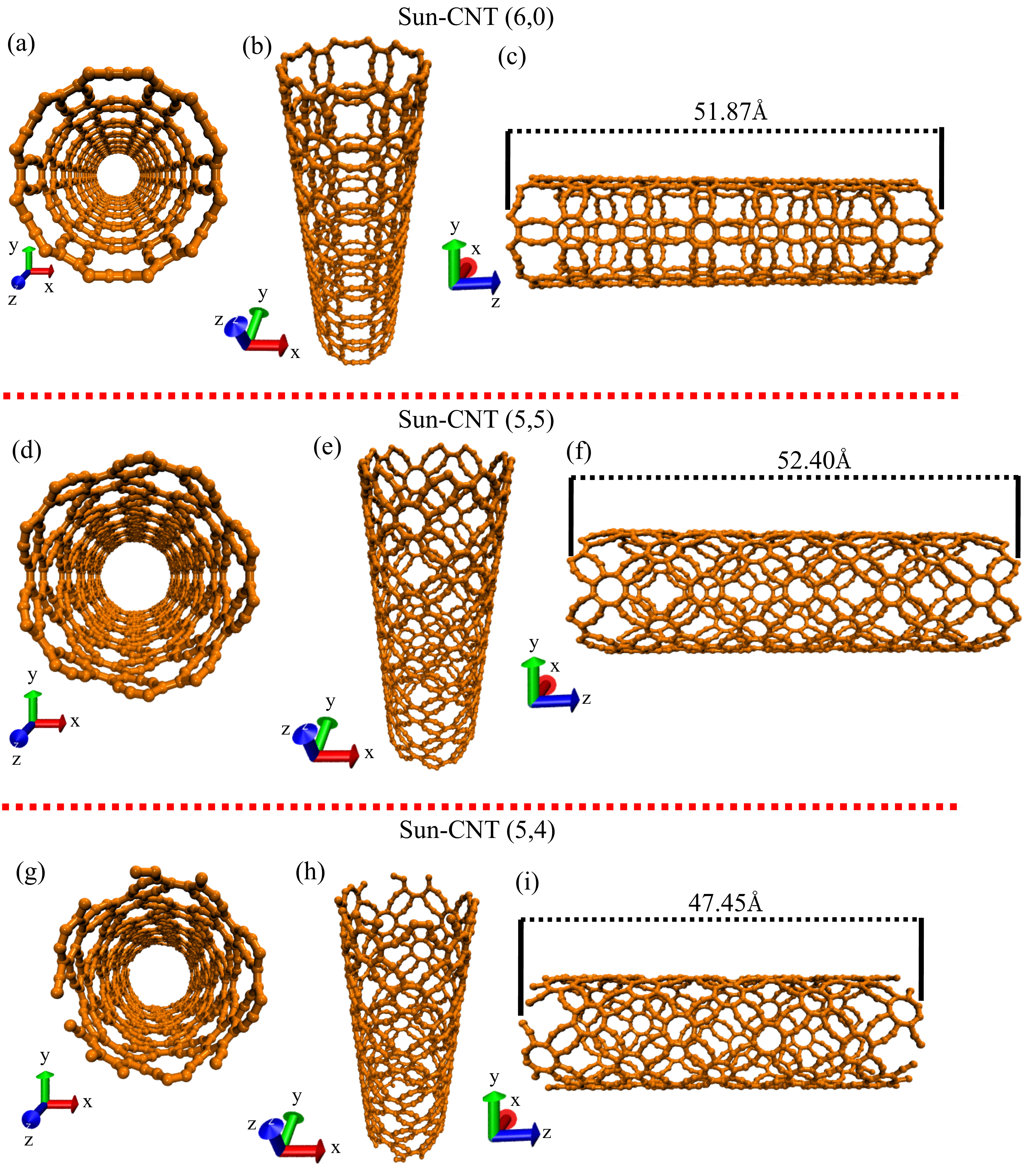}
\caption{\footnotesize{Representative nanostructural fully atomic model of Sun-graphene-based nanotube (Sun-CNT) zigzag $(6,0)$, armchair $(5,5)$ and, chiral $(5,4)$. (a) Front view in the van der Waals and dynamics bonds representation of Sun-CNT $(6,0)$, (b) longitudinal view of nanotube and (c) the side view of the nanotube. (d) Front view in the van der Waals and dynamics bonds representation of Sun-CNT $(5,5)$, (e) longitudinal view of nanotube and (f) the side view of the nanotube. (g) Front view in the van der Waals and dynamics bonds representation of sun-CNT $(5,4)$, (h) longitudinal view of nanotube and (i) the side view of the nanotube. Information on the radius, number of atoms, diameter and length of all Sun-CNTs studied in this research work can be viewed in the  table \ref{tab:Irida-CNT;Sun-CNT}.}}
\label{FIG:Sun-CNT:02}
\end{center}
\end{figure}

\section{Computational methodology}

Here we present the computational methodology procedures developed in this research work to estimate the elasticity of Irida-CNT and Sun-CNT (see table \ref{tab:Irida-CNT;Sun-CNT}). During the stretching process of the nanotubes studied in this research, using the open-source LAMMPs code ( Large-scale Atomic/Molecular Massively Parallel Simulator) \cite{plimpton1995fast}, the trajectories of the atomic positions are described by the classical molecular calculations with a focus on the physical/chemical properties of materials and nanostructures modeling. In the method of classical molecular dynamics (CMD), atoms are considered as classical particles, where the atomic interactions are described by classical interatomic force fields.

Thus, in this work, we used a classic reactive interatomic potential called ReaxFF \cite{van2001reaxff,mueller2010development}. This force field describes the atomic interaction and chemical bonds in Irida-CNTs and Sun-CNTs. ReaxFF is a modern force field that allows the breaking and formation of new chemical bonds during fully atomistic classical molecular dynamics simulations which was developed by Adri van Duin, William Andrew Goddard III and collaborators \cite{van2001reaxff}. ReaxFF is designed to be a force field close to classical and quantum chemical methods. Atomic parametrizations are developed with first-principle calculations and experimental results \cite{van2001reaxff}. ReaxFF is described in the bond order formalism along with descriptions of polarizable charges in the atomic description of reactive and non-reactive interactions. So, ReaxFF to accurately model both covalent and electrostatic interactions. Thus, the interatomic reactive force field is used in the description of physical/chemical properties of a diverse range of materials and nanostructures \cite{van2003reaxffsio,strachan2003shock,strachan2005thermal,senftle2016reaxff}. The summary of the energy contributions to the ReaxFF potential is as follows: \cite{van2001reaxff}:
\begin{eqnarray}
\label{esys}
E_{system}&=&E_{bond}+E_{over}+E_{angle}+E_{tors}\nonumber \\
&+&E_{vdWaals}+E_{Coulomb}+E_{Specific}.
\label{eq:energy-reax}
\end{eqnarray}
The ReaxFF potential includes energy contributions such as $E_{bond}$, which depends on the distance between atoms, and $E_{angle}$ and $E_{tors}$, which represent the energies related to valence angle strain and torsional angle strain respectively.
$E_{over}$ serves as a penalty to prevent atoms from being over-coordinated. $E_{Coulomb}$ and $E_{vdWaals}$ are energy contributions that account for electrostatic and dispersive interactions between all atoms. $E_{Specific}$ encompasses specific terms related to the system, including lone-pair interactions, conjugation effects, hydrogen bonding, and corrections associated with the $C_{2}$ parameter
\cite{van2001reaxff}. ReaxFF is divided into two types of energy contributions: bond-order-dependent and bond-order-independent. The bond order is determined by the interatomic distance.
\begin{eqnarray}
BO_{ij} &=& exp\left[ p_{bo,1} \cdot \left( \frac{r_{ij}}{r_{o}} \right)^{p_{bo,2}} \right] + exp\left[ p_{bo,3} \cdot \left( \frac{r_{ij}^{\pi}}{r_{o}} \right)^{p_{bo,4}} \right] + \nonumber\\ 
&+& exp\left[ p_{bo,5} \cdot \left( \frac{r_{ij}^{\pi \pi}}{r_{o}} \right)^{p_{bo,6}} \right], 
\label{eq:bond-order}
\end{eqnarray}
The bond order between atoms $i$ and $j$, denoted as $BO_{ij}$, determines the contribution of the interatomic distance to the atomic configurations in ReaxFF. This bond order is influenced by the interatomic distances of various types of bonds, including the $\sigma$ bond ($p_{bo,1}$ and $p_{bo,2}$), the first $\pi$ bond ($p_{bo,3}$ and $p_{bo,4}$), and the $\pi$-$\pi$ bond ($p_{bo,5}$ and $p_{bo,6}$). The interatomic distances for these bonds are dependent on the bond types, such as the $C - C$ bond ($\sigma$ bond with a distance of approximately 1.5 Å), the $\pi$ bond (with a distance of approximately 1.2 Å), and the $\pi$-$\pi$ bond (with a distance of approximately 1.0 Å). 

The fully atomistic visualizations, as well as the frames of the snapshots presented, the results of the stress/strain dynamics of all the Irida-CNTs and Sun-CNTs obtained by Visual Molecular Dynamics (VMD) \cite{humphrey1996vmd}. All CMD computational simulations of Irida-CNT and Sun-CNT stretching are performed at $300$ K, with periodic boundary conditions along with the $z$-axis. Table \ref{tab:Irida-CNT;Sun-CNT} illustrates the chirality, number of atoms, radius, diameter and length of the Irida-CNT and Sun-CNT. All computer simulations are started by performing a nanotube minimization process before the stretching process. The energy minimization of all carbon nanotubes is performed by iteratively adjusting atom coordinates. The Stoermer-Verlet time integration algorithm velocity-Verlet \cite{martys1999velocity} is used as the time integrator of CMD simulations performed by LAMMPS code. To adjust and balance the initial stress before the nanotube stretching process, we thermalize the nanotubes within the isothermal-isobaric set considering zero pressure in the nanotubes for $0.01$ ns \cite{evans1983isothermal}. Then, we performed a thermalization of the nanotubes for $0.01$ ns controlled by a Nose-Hoover \cite{hoover1985canonical} thermostat. In all CMD stretching simulations of the Irida-CNT and Sun-CNT, the temperature was kept constant at $300$K. The uniaxial stretching deformations of Irida-CNT and Sun-CNT were produced by increasing the simulation box size along the $z$-axis (periodic direction). The classical molecular dynamics simulations of stretching of all nanotubes were updated for each increment of $0.05$ fs. In the computational simulations, we used a constant engineering tensile strain rate $\zeta= 10^{-6}$ /fs. Length $L$ of Irida (Sun)-CNT evolves as $L(t) = L_{0}(1 + \zeta dt)$, where $L_{0}$ is the initial nanotube length and $dt$ is the elapsed time (in time units ($fs$)). The elasticity of Irida (Sun)-CNT was obtained by the physical quantity known as Young's modulus ($Y_{MOD}$), where its mathematical representation is given by ($Y_{MOD} = \frac{d \sigma_{zz}}{d \epsilon_{z}}$), where $\sigma_{zz}$ is
the component of the Virial stress tensor and $\epsilon_{z}$ is the deformation along
the uniaxial direction $z$-axis.

The stress tensor is defined as \cite{mcquarrie1987virial,de2016mechanical}:\\
\begin{eqnarray}
  \sigma_{\alpha \beta }=\frac{1}{V }\sum_{i}^{N}\left (m_{i}v_{\alpha i}v_{i\beta } + r_{i\alpha }f_{i\beta } \right ) ,
\label{Eq:08}
\end{eqnarray}
where, adapting to the variables of Eq. \ref{Eq:08} for uniaxial deformation in the $z$ direction, we then have mathematically \cite{mcquarrie1987virial,de2016mechanical}:\\:
\begin{eqnarray}
\sigma_{z} \equiv \sigma_{zz} = \frac{1}{V}\left[ \sum_{l=1}^{N} \left( m_{l}v_{lz}v_{zl} + r_{lz}F_{lz} \right) \right],
\label{Eq:08}
\end{eqnarray}
where $V$ is the volume of the hollow cylinder is defined as ($V = L_{0}\pi D_{CNT} h$). The $L_{0}$ is the initial Irida (Sun)-CNT length, $D_{CNT}$ is the diameter of nanotubes and $h = 3.35$\AA~ is the thickness value of carbon nanotubes-based Irida and Sun nanostructures. In Eq. \ref{Eq:08}, the variables $N$, $m$, $v$, $r$, and $f$ represent the total number of carbon atoms, the mass, velocity, position, and force per atom of each carbon atom in the Irida and Sun nanostructures made of carbon nanotubes that were investigated in this study. The von Mises stress tensor distribution per atom is defined as \cite{mcquarrie1987virial,de2016mechanical}: 
\begin{equation}
\sigma_\text{vonMises} = \sqrt{\left[\frac{(\sigma_{xx}-\sigma_{yy})^2 + (\sigma_{yy}-\sigma_{zz})^2 + (\sigma_{zz}-\sigma_{xx})^2 + 6(\tau_{xy}^2 + \tau_{yz}^2 + \tau_{zx}^2)}{2}\right]},
\label{Eq:09}
\end{equation}
The symbols $\sigma_{xx}$, $\sigma_{yy}$, and $\sigma_{zz}$ indicate the stresses in the $x$, $y$, and $z$ directions, respectively. The  $\tau_{xy}$, $\tau_{yz}$, and $\tau_{zx}$ represent the shear stress components. The computational simulations of stretch process evolution, the spatial atomic stress distribution average in Irida (Sun)-CNT, is calculated using the von Mises stress tensor (Eq. \ref{Eq:09}) \cite{mcquarrie1987virial,de2016mechanical}. The von Mises stress tensor is usually employed in the study of materials and nanostructures that are under mechanical load. Thus, the von Mises stress tensor shows us the regions with the highest concentration of mechanical stress. In the literature, we can show that the von Mises stress tensor is widely used in several areas of science, such as Chemistry \cite{an2011elucidation}, Biology \cite{lu2007inverse}, Engineering \cite{zienkiewicz1969elasto}, Medicine \cite{mian2005laser}, Materials Science \cite{qin2019optimization}, Classical Molecular Dynamics \cite{goel2014molecular} and Density Functional Theory (DFT) dynamics \cite{fago2004density}.

\subsection{Calculation of Poisson's ratio of the Irida-graphene-based (Irida-CNT) and Sun-graphene-based (Sun-CNT) nanotubes.}

The tubular shape of carbon nanotubes studied in this research article (Irida-CNT and Sun-CNT) the one-dimensional nanostructure subjected to a prescribed uniaxial deformation $\epsilon$ without the radial constraint, the evaluated the Poisson's ratio is given by \cite{wang2005size}:

\begin{eqnarray}
    \nu = - lim _{\epsilon = 0} \left( \frac{\epsilon^{*}}{\epsilon} \right),
\end{eqnarray}
where, $\nu$ Poisson's ratio, $\epsilon^{*}$ radial strain and $\epsilon$ uniaxial strain applied in the nanostructures. Thus, the Poisson's ratio, $\nu$, of Irida-CNT  - $(7,0)$ and $(4,4)$ and Sun-CNT - $(6,0)$, $(5,5)$ and $(5,4)$ was calculated by follows mathematical expression \cite{wang2005size}:\\

\begin{eqnarray}
    \nu = - \frac{\epsilon _{R}}{\epsilon _{z}},
\end{eqnarray}
The radial strain ($\epsilon _{R}$) and uniaxial strain ($\epsilon _{z}$) are defined in terms of deformation. The uniaxial strain can be directly measured from the rate of deformation in carbon nanotubes, while calculating the radial strain requires additional calculations. The length ($L$) of Irida-CNT and Sun-CNT is divided into equal-size slabs with thickness ($\Delta L = \frac{L}{n}$). The uniaxial strain is then calculated using the following mathematical expression:
\begin{eqnarray}
\epsilon_{R} = \frac{R^{(average)}_{\epsilon_{z}} - R^{(average)}_{0}}{R^{(average)}_{0}},
\end{eqnarray}
where, $R^{(average)}_{\epsilon_{z}}$ is the average Irida-CNT and Sun-CNT radius at the uniaxial strain $\epsilon_{z}$. $R^{(average)}_{0}$ is the average nanotube radius at the equilibrium, $\epsilon_{z}=0$, and is calculated by follows mathematical expression:
\begin{eqnarray}
R^{(average)}_{\epsilon_{z}} = n^{-1}\sum_{i=1}^{n} R^{(slab)}_{i}|_{\epsilon_{z}}.
\label{Eq:14}
\end{eqnarray}

In Eq. \ref{Eq:14}, $R^{(slab)}_{i}|_{\epsilon_{z}}$ is the average radius of the {\textit{i}}-esim circular slab of the Irida-CNT and Sun-CNT at uniaxial strain $\epsilon_{z}$. $R^{(slab)}_{i}$ is the average of the distance of each slab (set carbon atoms) to its center of mass. Is calculated by follows mathematical expression:
\begin{eqnarray}
R^{(slab)}_{i} = M^{-1}\sum_{\zeta 
= 1}^{N} r_{\zeta},
\label{Eq:15}
\end{eqnarray}
where, $r_{\zeta}$ is given by the mathematical expression:
\begin{eqnarray}
r_{\zeta} = \sqrt{\left( x_{\eta} - x_{CM}  \right)^{2} + \left( y_{\eta} - y_{CM}  \right)^{2}}.
\label{Eq:16}
\end{eqnarray}
In the Eq. \ref{Eq:16}, $x_{\eta}$ and $y_{\eta}$ are the planar coordinates of the $\eta$-esim carbon atoms ($M$ carbon atoms per slab). $x_{CM}$ and $y_{CM}$ are coordinates of the mass center of each slab of Irida-CNT and Sun-CNT. In the calculations to obtain Poisson's ratio, we consider $\Delta L = 1.5$\AA~. There is reasonable minimum number of atoms at each slab and also a reasonable number of different slabs in order to obtain a good average of the radius of the Irida-CNT and Sun-CNT \cite{brandao2023first}.

\section{Results and discussions}

\begin{figure}[ht!]
\begin{center}
\includegraphics[angle=0,scale=1.30]{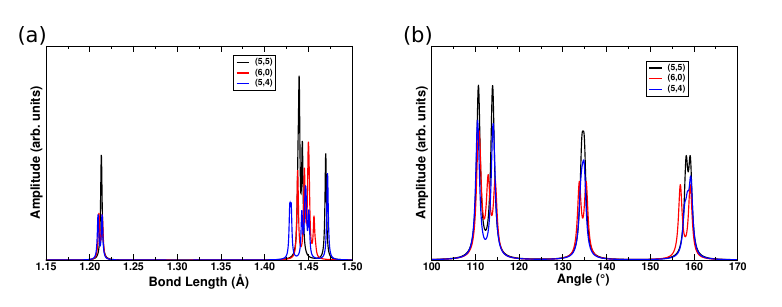}
\caption{\footnotesize{Bond length and bond angle distributions for Sun-CNT $(5,5)$, $(6,0)$ and $(5,4)$ . In (a) the bond length (\AA) and (b) the angle distributions, respectively. The results are for all simulated temperatures and normalized to the values at 10 K.}}
\label{FIG:Sun-CNT:angulos_bonds}
\end{center}
\end{figure}

Here, we represent the results of the fully atomistic reactive computational simulations based on the formalism described in section 3. Before uniaxial stretching mechanical load dynamics, all Irida-CNT and Sun-CNT nanotubes were minimized to correct the atomic geometry of their nanostructure. The results for the minimized nanotubes with the ReaxFF parameter set \cite{mueller2010development}, distributions of bond lengths and angles were obtained. Panels (a) and (b) in Figs. \ref{FIG:Irida-CNT(n,O):angulos_bonds} and \ref{FIG:Sun-CNT:angulos_bonds} show the different bond lengths and angle values which are distinguishable from the minimized nanotubes with ReaxFF set parameter $C - C$ bonds. In Figure \ref{FIG:Irida-CNT(n,O):angulos_bonds} (a), we present the bond length distribution analysis for two distinct Irida-Carbon Nanotubes (CNT) configurations, namely (4,4) and (7,0), depicted as black and red curves, respectively. Notably, for the Irida-CNT (4,4) structure, the bond lengths exhibit a distribution spanning the range of 1.406-1.439 \AA{}, with prominent peaks observed at 1.407, 1.421 and 1.436 \AA{}. These characteristics are indicative of the typical bond lengths found in pure $sp^{2}$  covalent carbon materials.

In contrast, the Irida-CNT (7,7) structure displays a broader distribution of bond lengths within the range of 1,420 - 1,444 \AA{}. Importantly, this broadening should not be construed as a loss of crystallinity; rather, it suggests a higher degree of variability in bond lengths within this region. This variability is attributed to an increased number of covalent bonds in comparison to the Irida-CNT (4,4) configuration. The bond length distribution peaks are situated at approximately 1.420, 1.430 and 1.440 \AA{}, mirroring the $sp^{2}$  carbon nature, similar to the Irida-CNT (4,4) structure.

An alternative method for characterizing these nanostructures involves the analysis of the distribution of angles formed by covalent bonds. As depicted in Fig. \ref{FIG:Irida-CNT(n,O):angulos_bonds} (b), both structural configurations present distinct peaks within the angular distribution profile, precisely situated at 60.05$^{\circ}$, 119.75$^{\circ}$, and 149.77$^{\circ}$. Notably, the first peak is linked to the inherent angle formed by atoms constituting the triangular conformation, while the second peak signifies atoms organized in hexagonal arrangements, and the third peak corresponds to atoms arranged in octagonal conformations.

\begin{figure}[ht!]
\begin{center}
\includegraphics[angle=0,scale=1.30]{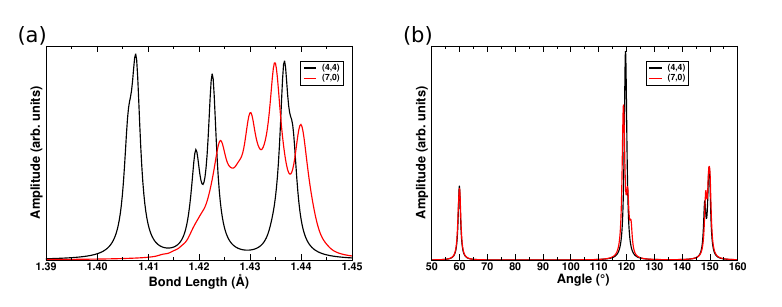}
\caption{\footnotesize{Bond length and bond angle distributions for Irida-CNT $(4,4)$ and $(7,0)$ . In (a) the bond length (\AA) and (b) the angle distributions, respectively. The results are for all simulated temperatures and normalized
to the values at 10 K.}}
\label{FIG:Irida-CNT(n,O):angulos_bonds}
\end{center}
\end{figure}

Fig. \ref{FIG:Sun-CNT:angulos_bonds} represents similar results for Sun-CNT $(5,5)$, $(6,0)$ and $(5,4)$. In Figure \ref{FIG:Sun-CNT:angulos_bonds} (a), we present the bond length distribution analysis for three distinct Sun-Carbon Nanotubes (CNT) configurations, namely (5,5), (6,0) and (5,4), depicted as black, red and blue curves, respectively. For all Sun-CNT structures, the bond lengths exhibit a distribution spanning the range of 1.211-1.459 \AA{}, with prominent peaks observed at 1.211, 1.431, 1.442 and 1.471 \AA{}. These characteristics are indicative of the typical bond lengths found in pure $sp^{2}$  covalent carbon materials.

As depicted in Figure \ref{FIG:Sun-CNT:angulos_bonds} (b), the three structural configurations present distinct peaks within the angular distribution profile, varying for 110.57$^{\circ}$ to 159.24$^{\circ}$, with peaks situated about 112.34$^{\circ}$, 119.75$^{\circ}$, and 149.77$^{\circ}$.

In Figs. \ref{FIG:Irida-CNT(n,O);Stress/Strain:03} -\ref{FIG:Sun-CNT(n,m);Stress/Strain:07}, we represented stress/strain curves obtained by uniaxial strain applied in $z$-direction for all carbon nanotubes of Irida-CNT and Sun-CNT. The results from the reactive of CMD simulations with interatomic force field ReaxFF \cite{mueller2010development} are obtained for the stress/strain curves of Irida-CNT like-zigzag $(5,0)$, $(6,0)$, $(7,0)$, $(8,0)$, $(9,0)$ and $(10,0)$, like-armchair $(3,3)$, $(4,4)$, $(5,5)$, $(6,6)$, $(7,7)$ and $(8,8)$, as well, Sun-CNT like-zigzag $(4,0)$, $(5,0)$, $(6,0)$, $(7,0)$, $(8,0)$ and $(9,0)$, like-armchair $(3,3)$, $(4,4)$, $(5,5)$, $(6,6)$, $(7,7)$ and $(8,8)$ and like-chiral $(3,2)$, $(4,3))$, $(5,4)$, $(6,5)$, $(7,6)$ and $(8,7)$ at room temperature. These results confirm the fact that Irida-CNTs and Sun-CNTs have chirality-dependent mechanical properties under uniaxial tensile strength. It is found that the stress/strain curves show different mechanical responses for the Irida-CNTs and Sun-CNTs nanotubes.
The Irida-CNTs are similar to the ones observed in brittle materials and/or nanostructures under load strain while Sun-CNTs are similar to ductile materials and/or nanostructures.  It is worth pointing out that Irida-CNTs are more flexible than Sun-CNTs. This aspect indicates that Irida-CNTs would behave as flexible, high-strength nanostructures. 

Nevertheless, from the stress/strain curves are predicted the Young’s modulus for all Irida-CNTs and Sun-CNTs investigated in the theoretical research work. All results obtained from Young's Modulus by reactive CMD are shown in Table \ref{Table:Values:Results:01:Irida-CNT;Sun-CNT}. The stress-strain curves for Irida-CNT like-zigzag and like-armchair (Figs. \ref{FIG:Irida-CNT(n,O);Stress/Strain:03} and \ref{FIG:Irida-CNT(n,O);Stress/Strain:03a}), respectively. The relationship between the curves (stress (GPa) vs. strain (\%)) is characterized by the existence of a linear (elastic) regime and an unusual configuration on the curve, where we have abrupt drops in stress, originated from the nanostructural reconstruction (see Fig. \ref{FIG:Irida-CNT(n,0)(n,n);snapshot_reconstruction_nanostructural:09}). For Sun-CNTs like-zigzag, like-armchair and like-chiral we can see linear (elastic) and plastic regimes (see Fig. \ref{FIG:Sun-CNT(n,O);Stress/Strain:05}, \ref{FIG:Sun-CNT(n,n);Stress/Strain:06} and \ref{FIG:Sun-CNT(n,m);Stress/Strain:07}). The bonds start to break until they reach a complete fracture, which is characterized by an abrupt stress drop. The values of Young’s modulus for Irida-CNTs are slightly larger range ($640.34$ GPa  -  $825.00$ GPa) and UTS range ($93.33$ GPa - $109.11$ GPa ) to strained configuration, when compared to Sun-CNTs. Also, the large flexibility of Irida-CNTs, where the critical strain estimates quantity values of Irida-CNTs (range from $14$\% up to $28.41$\%) are approximately larger than the corresponding ones of Sun-CNTs ($18.50$\% up to $27.38$\%). For Sun-CNTs, the Young's Modulus range ($200.44$ GPa - $472.84$ GPa) and UTS range ($73.68$ GPa - $90.48$ GPa). Comparing the values of Young's modulus for Irida-CNTs and Sun-CNTs with the values for conventional carbon nanotubes CNTs found in the literature obtained by reactive (ReaxFF) CMD and calculations based on first principles (DFT) values ranging from ($821$ GPa - $995$ GPa) \cite{de2019elastic} and values of $1020$ GPa obtained by Krishnan et al. ($1998$) \cite{krishnan1998young}. With AIREBO interatomic potential, obtained by Michał, B. et {\textit{al.}} (2012), of the $910$ GPa \cite{michal2012mechanical}. Finally, comparing the quantities of elastic modulus between Irida-CNT, Sun-CNT and CNT, we observed in our results that Irida-graphene and Sun-graphene have a lower Young's Modulus.

\begin{figure}[ht!]
\begin{center}
\includegraphics[angle=0,scale=0.28]{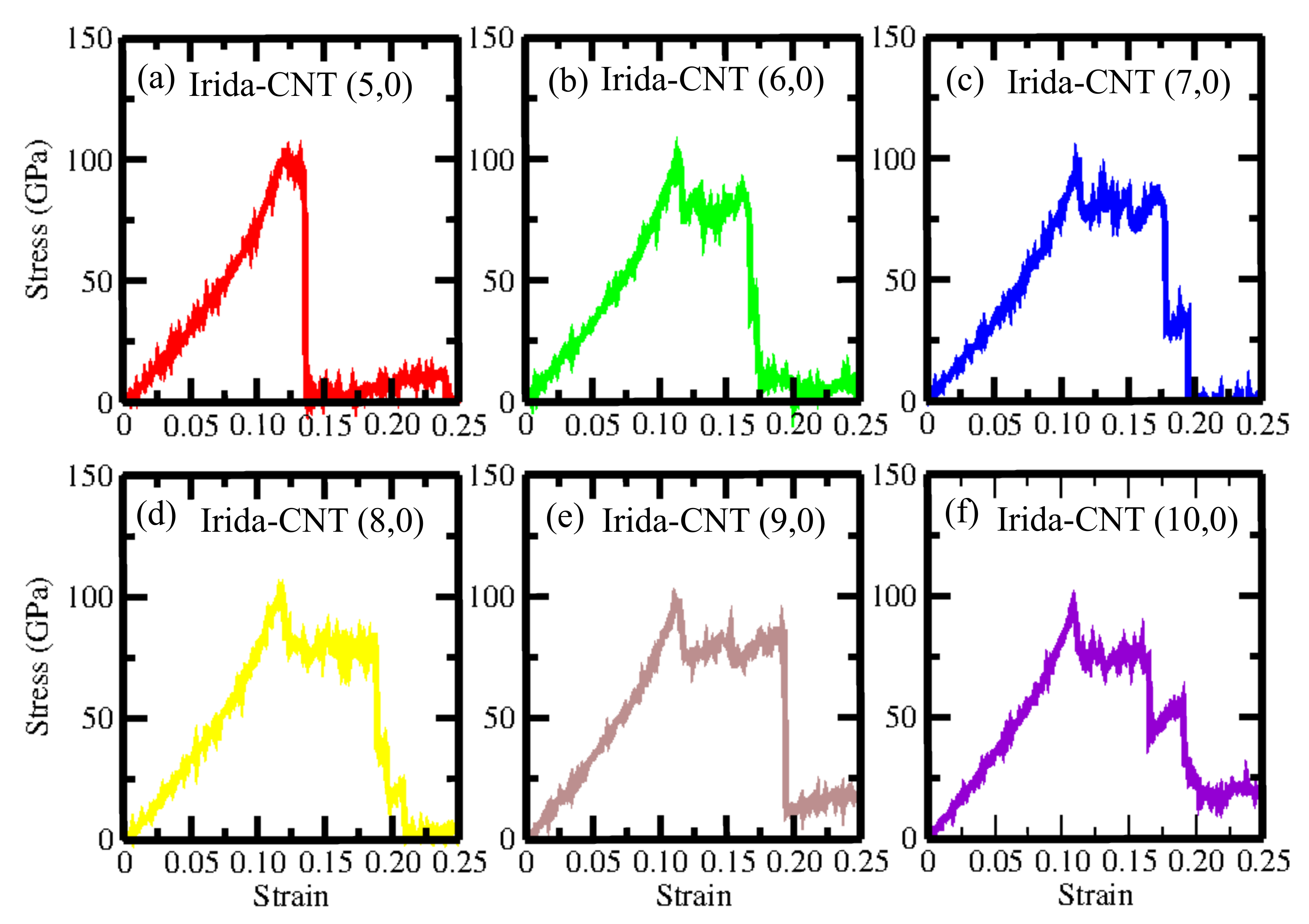}
\caption{\footnotesize{Stress versus strain curves for Irida-CNT $(n,0)$ predicted by reactive classical molecular dynamics simulations with the ReaxFF interatomic potential at $300 K$ with uniaxial strain applied in the $Z$ direction.}}
\label{FIG:Irida-CNT(n,O);Stress/Strain:03}
\end{center}
\end{figure}
\begin{figure}[ht!]
\begin{center}
\includegraphics[angle=0,scale=0.28]{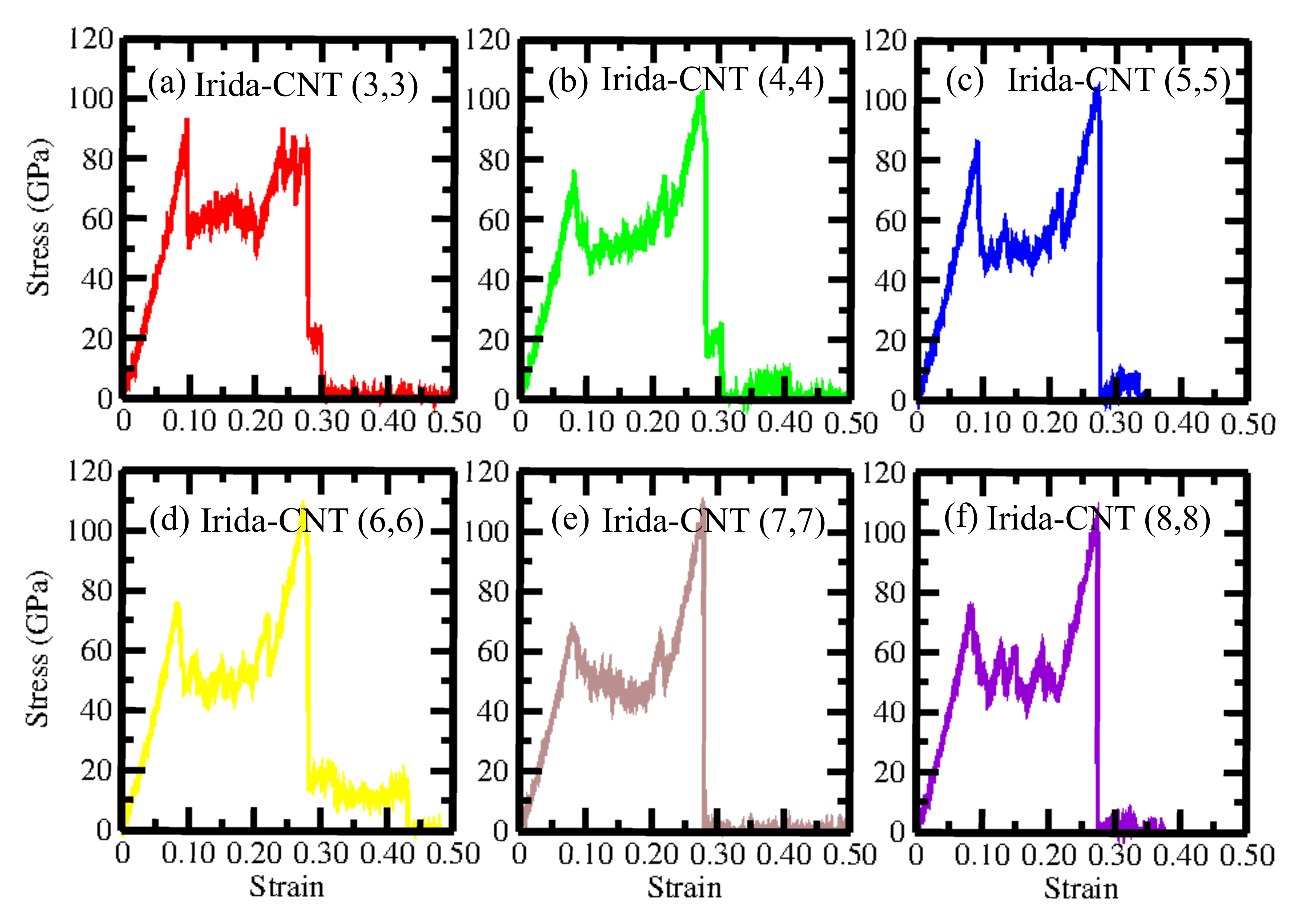}
\caption{\footnotesize{Stress versus strain curves for Irida-CNT $(n,n)$ predicted by reactive classical molecular dynamics simulations with the ReaxFF interatomic potential at $300 K$ with uniaxial strain applied in the $Z$ direction.}}
\label{FIG:Irida-CNT(n,O);Stress/Strain:03a}
\end{center}
\end{figure}
\begin{figure}[ht!]
\begin{center}
\includegraphics[angle=0,scale=0.28]{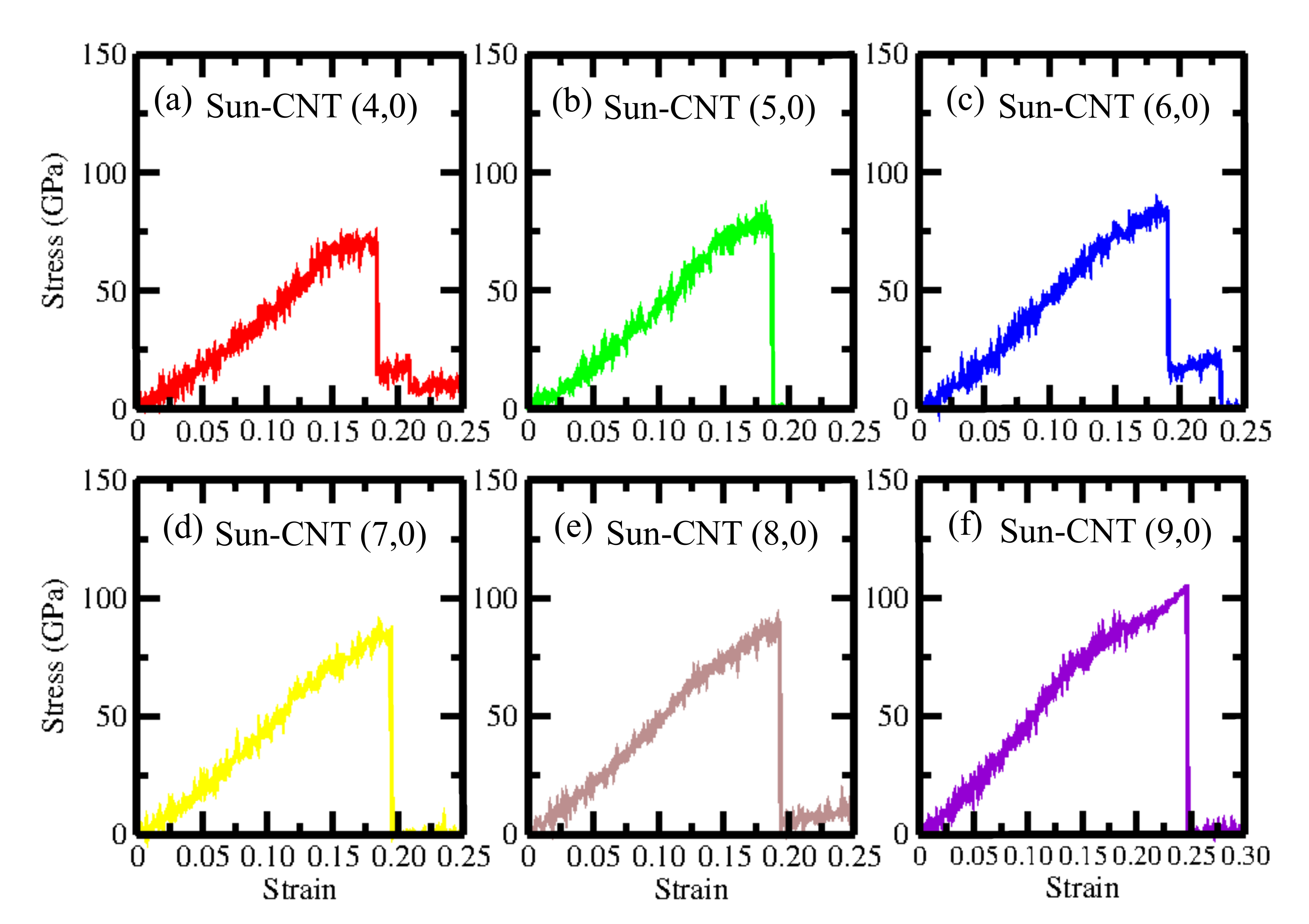}
\caption{\footnotesize{Stress versus strain curves for Sun-CNT $(n,0)$ predicted by reactive classical molecular dynamics simulations with the ReaxFF interatomic potential at $300 K$ with uniaxial strain applied in the $Z$ direction.}}
\label{FIG:Sun-CNT(n,O);Stress/Strain:05}
\end{center}
\end{figure}
\begin{figure}[ht!]
\begin{center}
\includegraphics[angle=0,scale=0.28]{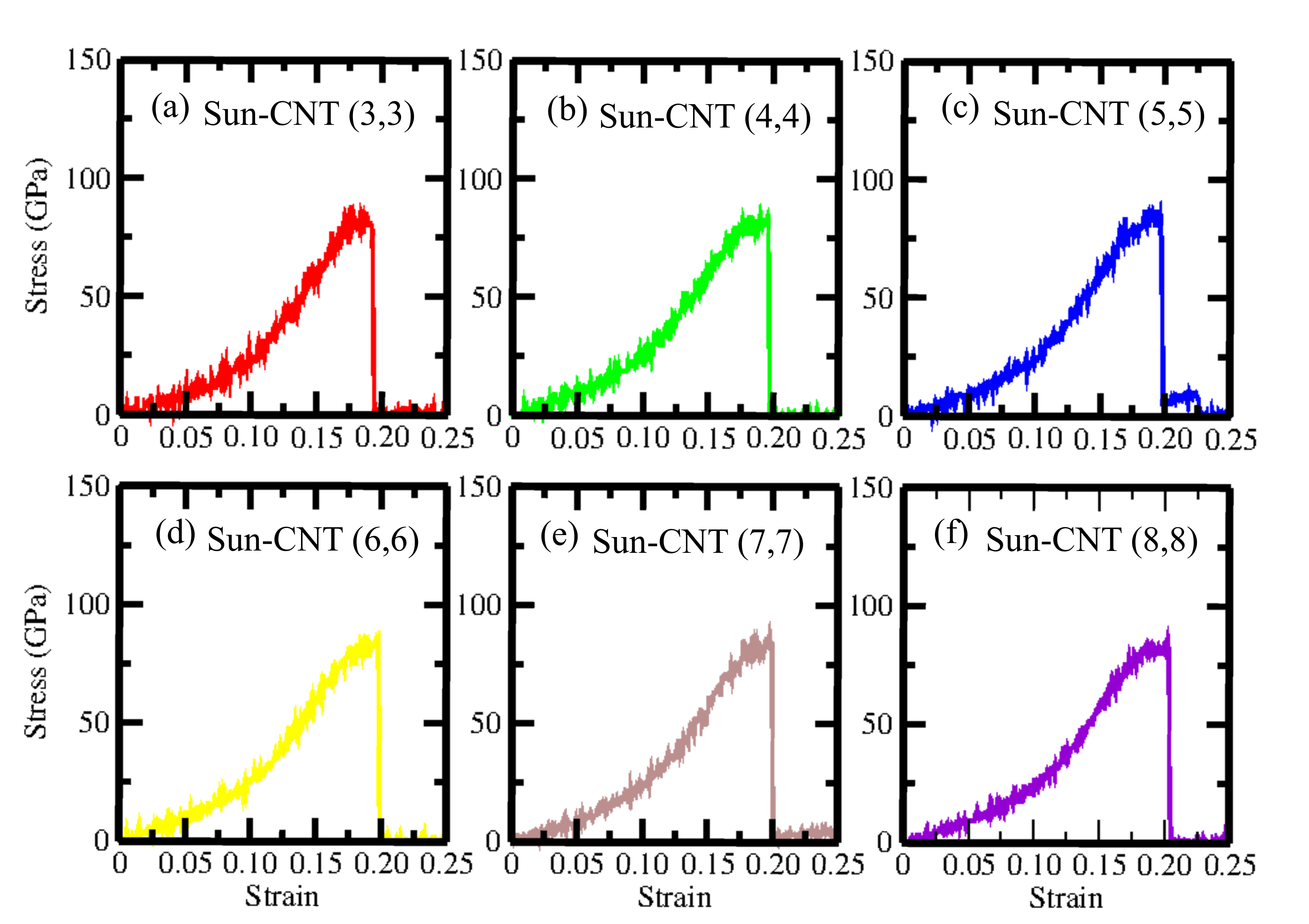}
\caption{\footnotesize{Stress versus strain curves for Sun-CNT $(n,n)$ predicted by reactive classical molecular dynamics simulations with the ReaxFF interatomic potential at $300 K$ with uniaxial strain applied in the $Z$ direction.}}
\label{FIG:Sun-CNT(n,n);Stress/Strain:06}
\end{center}
\end{figure}
\begin{figure}[ht!]
\begin{center}
\includegraphics[angle=0,scale=0.28]{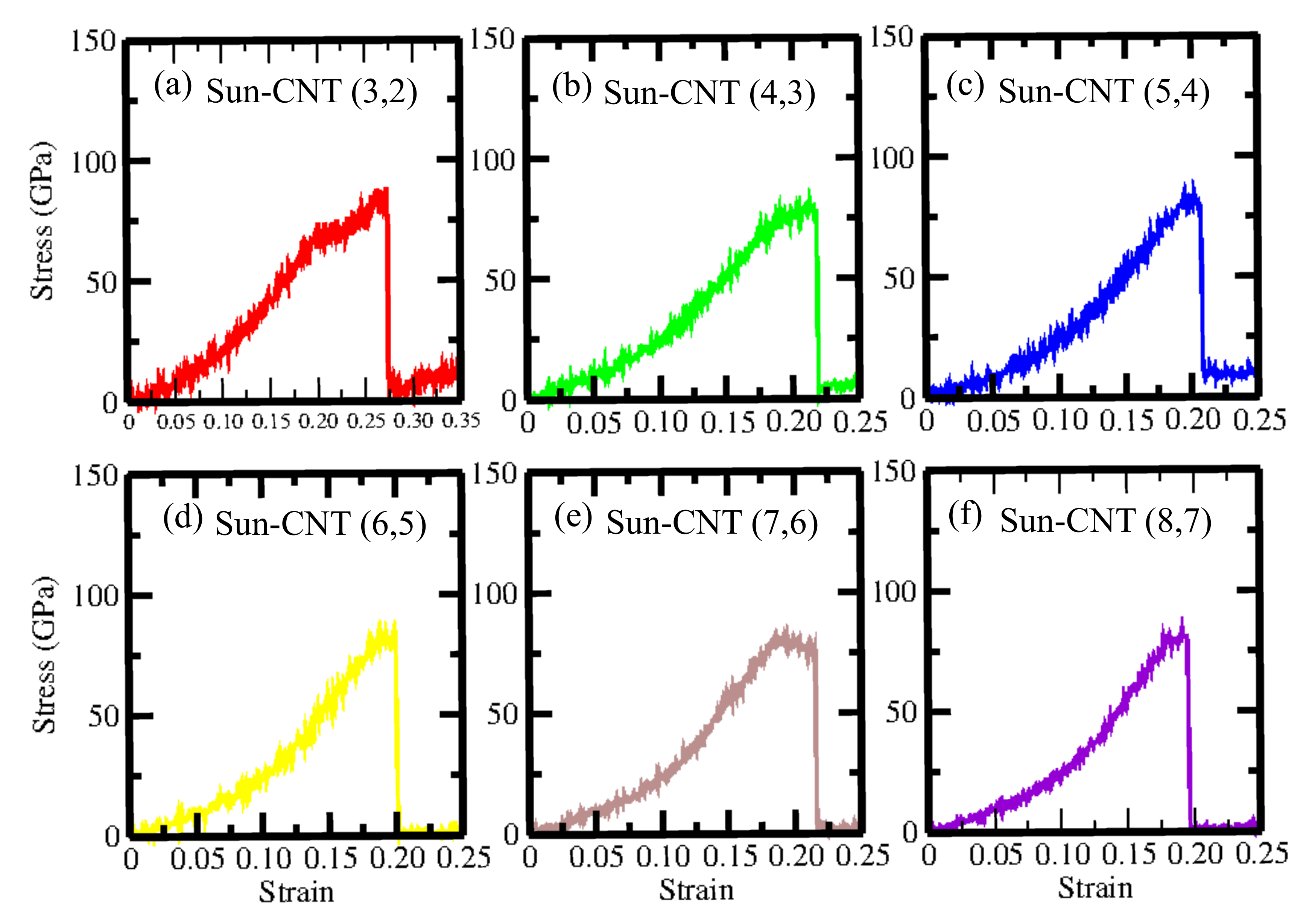}
\caption{\footnotesize{Stress versus strain curves for Sun-CNT $(n,m)$ predicted by reactive classical molecular dynamics simulations with the ReaxFF interatomic potential at $300 K$ with uniaxial strain applied in the $Z$ direction.}}
\label{FIG:Sun-CNT(n,m);Stress/Strain:07}
\end{center}
\end{figure}

In addition, our results show an unusual behavior in the Irida-CNTs, where there is a sudden drop in stress before the complete fracture of the Irida-CNTs. This is due to nanostructural reconstructions, wherein in the stretching regimen some bonds are broken (chemical bonds aligned in the direction of uniaxial stretching in the $z$-direction), The Irida-CNT shows a nanostructural atomistic reconstruction transition and so with increasing stretching, until the complete rupture of the Irida-CNT configured by the separation of the nanotube in two parts. To better explain the sudden drop in stress before complete fracture, see Fig.\ref{FIG:Irida-CNT(n,0)(n,n);snapshot_reconstruction_nanostructural:09} illustrates the sudden drop process in stress before the complete fracture. In this figure, we showed the representative reactive CMD snapshots of a tensile stretch atomic reconstruction of zigzag Irida-CNTs $(7,0)$ and armchair $(4,4)$.

\begin{figure}[ht!]
\begin{center}
\includegraphics[angle=0,scale=0.18]{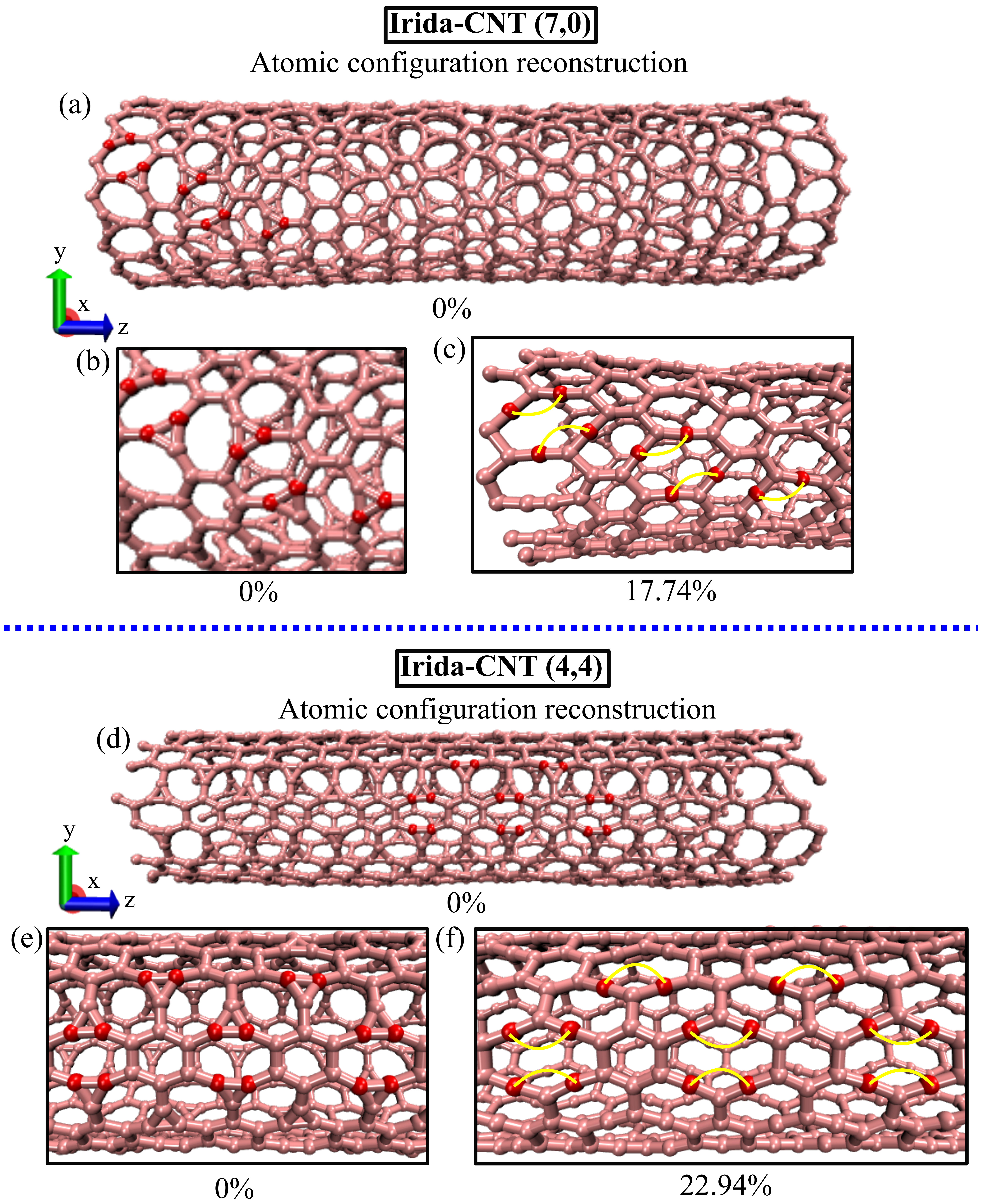}
\caption{\footnotesize{Representative reactive CMD snapshots of a tensile stretch atomic reconstruction of zigzag Irida-CNTs  $(7,0)$ and, armchair $(4,4)$. (a) Irida-CNT $(7,0)$ at $0$\% of strain with atoms selected in red color, in (b) a zoomed view of selected carbon atoms in the zigzag Irida-CNTs $(7,0)$ in red color at $0$\% of strain, (c) the break of chemical bonds of labeled on carbon atoms in red color, showing a nanostructural atomic reconstruction at $17.74$\% of strain. In (d) Irida-CNT $(4,4)$ at $0$\% of strain with atoms selected in red color. In (e) a zoomed view of selected carbon atoms in the zigzag Irida-CNTs $(4,4)$ in red color at $0$\% of strain and, (f) the break of chemical bonds of labeled carbon atoms in red color, showing a nanostructural atomic reconstruction at $22.94$\% of strain.}}
\label{FIG:Irida-CNT(n,0)(n,n);snapshot_reconstruction_nanostructural:09}
\end{center}
\end{figure}

In panels (a) and (b) in Fig.\ref{FIG:Irida-CNT(n,0)(n,n);snapshot_reconstruction_nanostructural:09}, we highlighted the carbon atoms in red, where during the mechanical load process of uniaxial stretching of the Irida-CNT, the bonds in the highlighted atoms are broken. It should be noted that without loss of the configuring in the mechanical fracture of the nanotube, we have a nanostructural reconstruction that can be better visualized in panel (b), where the lines in yellow color represent the breaking of the chemical bonds $C - C$, where the Irida-CNT undergoes a nanostructural transition before the complete mechanical fracture. This explains why Irida-CNTs support a higher uniaxial strain when compared to Sun-CNTs.

\begin{figure}[ht!]
\begin{center}
\includegraphics[angle=0,scale=0.21]{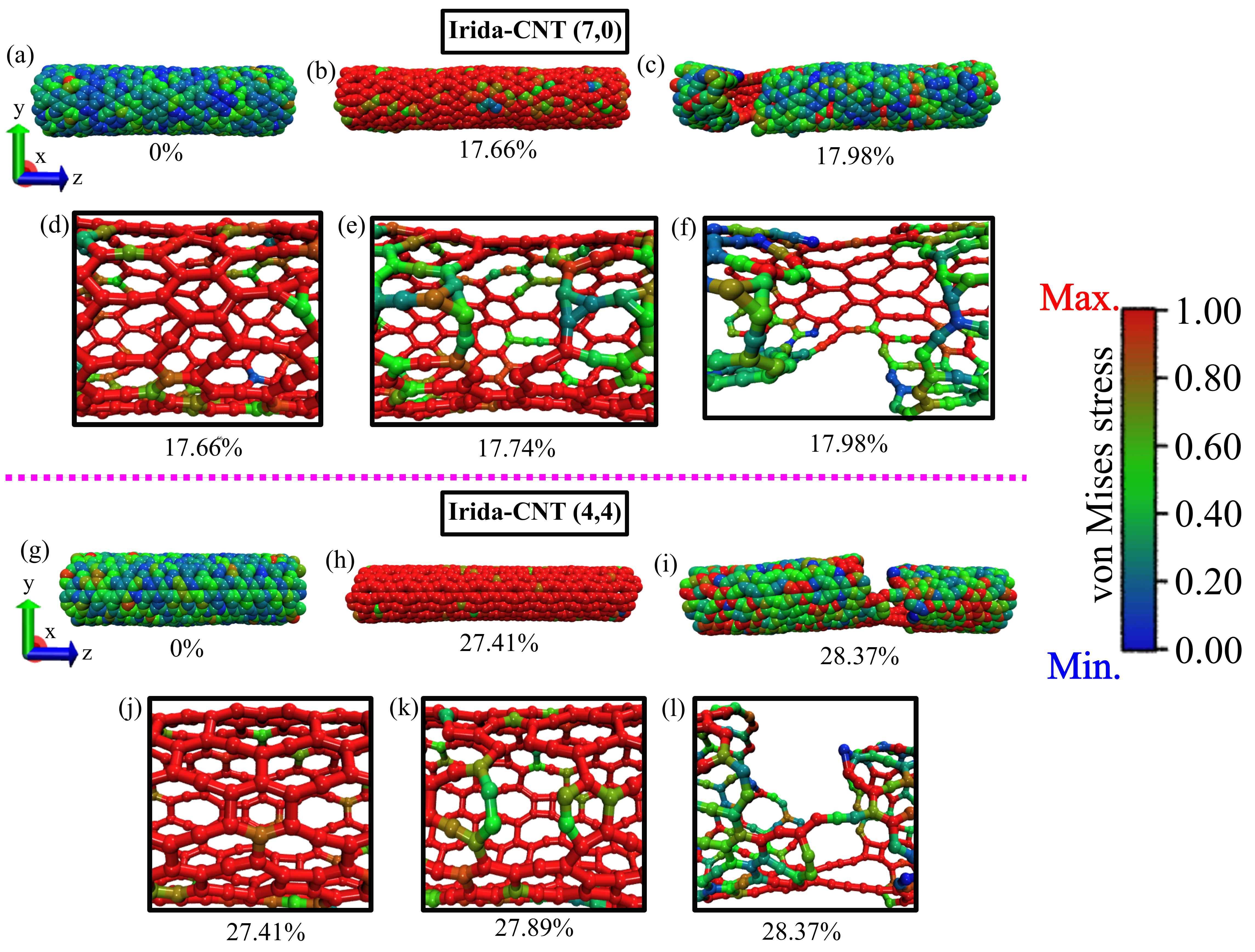}
\caption{\footnotesize{Representative reactive CMD snapshots of a tensile stretch of zigzag Irida-CNTs $(7,0)$ and, armchair $(4,4)$. (a) Irida-CNT $(7,0)$ at $0$\% of strain, in (b) the Irida-CNT strained at $17.66$\% of strain, (c) Irida-CNT $(7,0)$ completely fractured at $17.98$\% of strain. In (d), (e) and (f) a zoomed view of the starting of bond breaking Irida-CNT $(7,0)$. In (g) Irida-CNT $(4,4)$ at $0$\% of strain, in (h) the Irida-CNT strained at $27.41$\% of strain and (i) Irida-CNT $(4,4)$ completely fractured at $28.37$\% of strain. The boxes below, (j), (k) and (l)  a zoomed view of the starting of bond breaking  Irida-CNT $(4,4)$. The lateral color bar showed a view of the strained nanotube colored according to the von Mises stress values (low stress in blue and high stress in red)}}
\label{FIG:Irida-CNT(n,0)(n,n);snapshot:08}
\end{center}
\end{figure}

In Fig. \ref{FIG:Irida-CNT(n,0)(n,n);snapshot:08}, The Irida-CNT and Sun-CNT nanostructural failure processes can be better understood following the evolution of the von Mises stress distributions from the reactive CMD snapshots of the tensile stretch (see Figs. \ref{FIG:Irida-CNT(n,0)(n,n);snapshot:08} and \ref{FIG:Sun-CNT(n,0)(n,n)(n,m);snapshots}). 
\begin{figure}[ht!]
\begin{center}
\includegraphics[angle=0,scale=0.21]{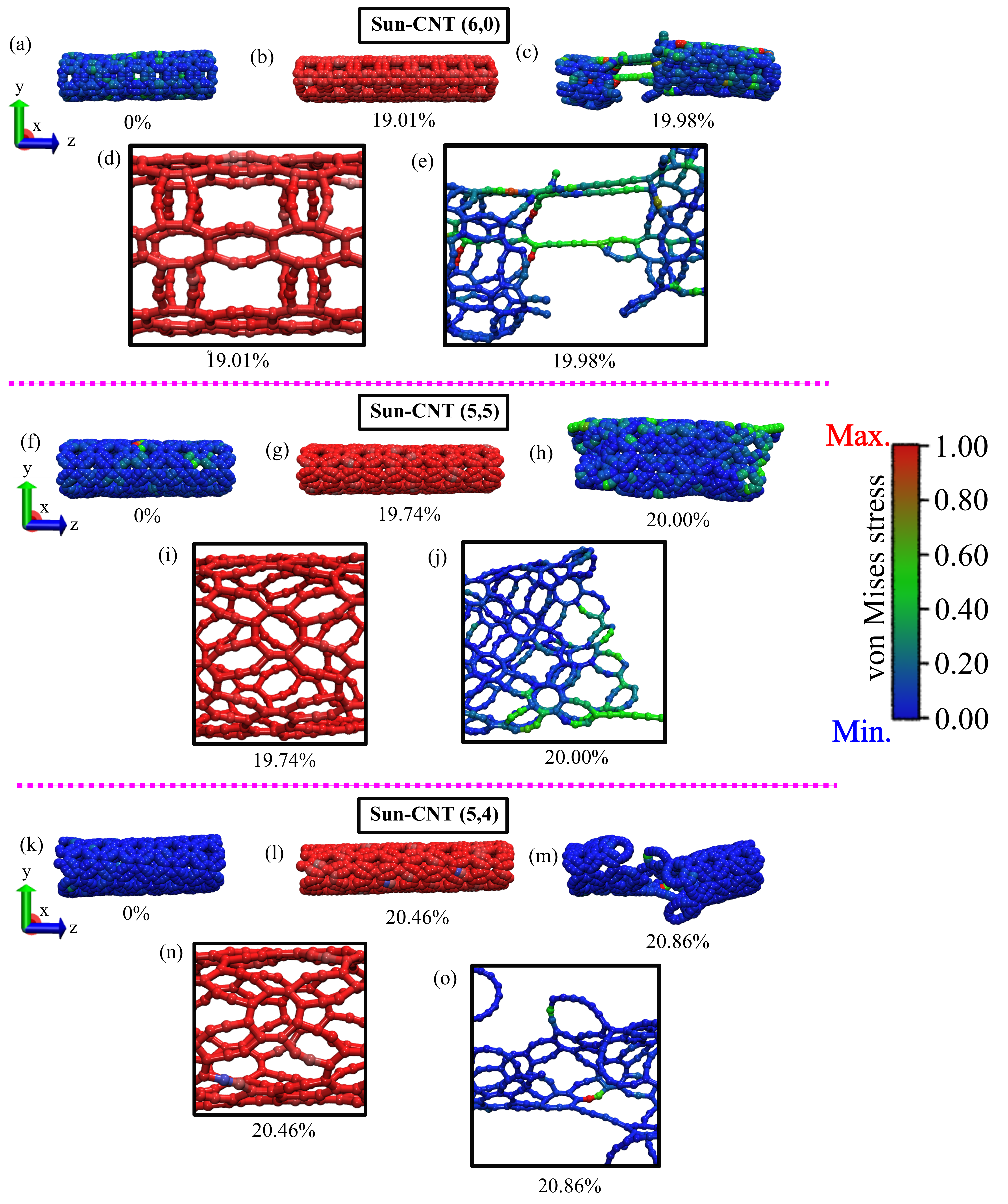}
\caption{\footnotesize{Representative reactive CMD snapshots of a tensile stretch of zigzag Sun-CNTs $(6,0)$, armchair $(5,5)$ and chiral $(5,4)$. (a) Sun-CNT $(6,0)$ at $0$\% of strain, in (b) the Sun-CNT strained at $19.01$\% of strain, (c) Sun-CNT $(6,0)$ completely fractured at $19.98$\% of strain. In (d) and (e) a zoomed view of the starting of bond breaking Sun-CNT $(6,0)$. In (f) Sun-CNT $(5,5)$ at $0$\% of strain, in (g) the Sun-CNT strained at $19.74$\% of strain and (h) Sun-CNT $(5,5)$ completely fractured at $20.00$\% of strain. The boxes below, (i) and (j) a zoomed view of the starting of bond breaking  Sun-CNT $(5,5)$. In (k) Sun-CNT $(5,4)$ at $0$\% of strain, in (l) the Sun-CNT strained at $20.46$\% of strain and (m) Sun-CNT $(5,4)$ completely fractured at $20.86$\% of strain. The boxes below, (n) and (o) a zoomed view of the starting of bond breaking  Sun-CNT $(5,4)$. The lateral color bar showed a view of the strained nanotube colored according to the von Mises stress values (low stress in blue and high stress in red)}}
\label{FIG:Sun-CNT(n,0)(n,n)(n,m);snapshots}
\end{center}
\end{figure}

Based on the theoretical results obtained by reactive CMD it is possible to observe high-stress accumulation lines (in red color) along the bonds parallel to the externally applied uniaxial strain in $z$-direction. In Fig. \ref{FIG:Irida-CNT(n,0)(n,n);snapshot:08}, we showed the nanofracture patterns for Irida-CNT $(7,0)$ and $(4,4)$. The broken bonds (Irida-CNT $(7,0)$) occur in the bonds arranged parallel to the applied uniaxial strain direction ($z$ - direction) in the Irida-CNTs, after atomic reconfiguration between carbon atoms (atomic reconstruction, see Fig. \ref{FIG:Irida-CNT(n,0)(n,n);snapshot_reconstruction_nanostructural:09}). 
Our results showed that the increase of the ring formed from $8$ atoms to $10$ carbon atoms, making the Irida-CNT more flexible. The initial failure begins in the single covalent bond in the ring composed of $10$ carbon atoms, after the nanostructural reconfiguration (reconstruction of the chemical bonds between the carbon atoms during stretching before the onset of mechanical failure). The initial morphological nanostructural shape of the nanopores in Irida-CNTs changes to a more stretched-like shape, after atomistic reconfiguration (see Fig. \ref{FIG:Irida-CNT(n,0)(n,n);snapshot:08} (d) at $17.66$\%~of uniaxial strain applied). The nanofracture initiation occurs at $17.74$\% of uniaxial strain rate (see Fig. \ref{FIG:Irida-CNT(n,0)(n,n);snapshot:08} (e)) and complete nanofracture at $17.78$\%~ of uniaxial stress, where the fully atomistic configuration is represented by the Irida-CNT separated into two parts. These evolution patterns are consequences of the high Irida-CNT flexibility, and it is responsible for the significant differences in the critical strain compared with Sun-CNTs. The same effect of mechanical failure is observed for Irida-CNT $(4,4)$, and thus for all Irida-graphene nanotubes invested in this theoretical research work. 

In Fig. \ref{FIG:Sun-CNT(n,0)(n,n)(n,m);snapshots}, we showed nanofracture patterns for Sun-CNT $(6,0)$, $(5,5)$ and $(5,4)$. In these theoretical results obtained by reactive CMD, we obtain mechanical nanofracture patterns, as well as different values of elastic quantities, justified by the different morphology between Irida-graphene and Sun-graphene nanotubes. For the Sun-CNTs, we can see high-stress accumulation lines (in red color) along the bonds parallel to the externally applied uniaxial strain in $z$-direction (see Fig. \ref{FIG:Sun-CNT(n,0)(n,n)(n,m);snapshots}). In Fig. \ref{FIG:Sun-CNT(n,0)(n,n)(n,m);snapshots} (a) up to (e), we showed the nanofracture process of Sun-CNT $(6,0)$. In Fig. \ref{FIG:Sun-CNT(n,0)(n,n)(n,m);snapshots} (a) the Sun-CNT at $0$\%~ of uniaxial strain, in (b) completely stretched to $19.01$\% of strain and (c) the nanotube completely nanofractured at $19.98$\%~ of strain. In the small boxes (d) ($19.01$\% of strain) and (e) ($19.98$\%~ of strain), a zoomed-in view of van der Walls and dynamic bonds view, fully stretched and nanofractured Sun-CNT $(6,0)$, respectively. The mechanical failure begins with the simple covalent bond (parallel to the direction of uniaxial deformation applied in the $z$-direction) in the ring formed by $8$ carbon atoms that connect to the other two groups of rings with the same atomistic configuration, with the formation of LACs. In (f) up to (j), we showed the mechanical failure processes for the Sun-CNT $(5,5)$. In (a) at null strain, (b) completely strained at $19.74$\%~ of strain and (h) complete nanofractured at $20.00$\%~ of strain. The small boxes showed a fully tensioned in the van der Walls and dynamics bonds view and complete nanofracture (begin of mechanical failure in single bonds aligned in the uniaxial strain in the $z$-direction), of the Sun-CNT $(5,5)$, respectively. In (k) up to (o), we showed the mechanical failure processes for the Sun-CNT $(5,4)$. In (k) the Sun-CNT $(5,4)$ at null strain, in (l) fully stretched at $20.46$\%~ of strain and (m) complete nanofractured at $20.86$\% of strain. The small boxes showed a fully tensioned in the van der Walls and dynamics bonds view and complete nanofracture (begin of mechanical failure in single bonds aligned in the uniaxial strain in the $z$-direction), of the Sun-CNT $(5,4)$, respectively (see Fig. \ref{FIG:Sun-CNT(n,0)(n,n)(n,m);snapshots}).  

\begin{figure}[ht!]
\begin{center}
\includegraphics[angle=0,scale=1.10]{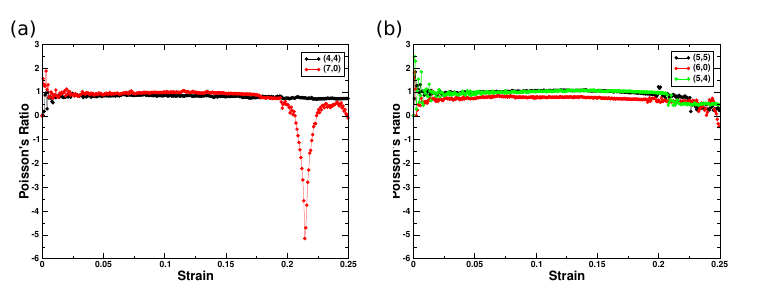}
\caption{\footnotesize{Graphic representation of Poisson's Ratio versus strain for (a) Irida-CNT $(4,4)$ and $(7,0)$ and (b) Sun-CNT $(5,5)$, $(6,0)$ and $(5,4)$ at room temperature.}}
\label{FIG:PR}
\end{center}
\end{figure}

Fig. \ref{FIG:PR}, shows the graphical Poisson’s coefficients representation of Irida-CNT $(7,0)$ and $(4,4)$ (Fig. \ref{FIG:PR} (a)) and Sun-CNT $(6,0)$, $(5,5)$ and $(5,4)$ (Fig. \ref{FIG:PR} (b)) as a function of the uniaxial strain applied in $z$-direction. As we can see in Fig. \ref{FIG:PR} (a), the Poisson ratio of Irida-CNT $(7,0)$ and $(4,4)$ as a 
function of the applied deformation in the order of 1,0 is the same for the Sun-CNT $(6,0)$, $(5,5)$ and $(5,4)$. After subjecting Irida-CNT $(7,0)$ to a 20\% strain, a noticeable decline in the Poisson's coefficient is observed, eventually reaching a remarkable negative value of -5.15. Conversely, Irida-CNT $(4,4)$ maintains a consistently positive Poisson's coefficient even under higher strain conditions. Regarding Sun-CNT, a positive Poisson's coefficient is consistently observed, except Sun-CNT $(6,0)$, which exhibits a negative Poisson's coefficient when subjected to strain levels approaching 25\%

The Poisson’s coefficients present us with a result of an elastic property of the nanotubes studied in this research article where it measures the transversal deformation of the Irida-CNTs and Sun-CNTs, with homogeneous and isotropic atomistic configurations. The graphical results showed that the values for the Poisson's ratio are different between Sun-CNT and Irida-CNTs. This result is consistent with the fact that Irida-CNT and Sun-CNT nanotubes have different nanostructural morphologies. For the Irida-CNTs, we can see an atomistic configuration densely packed between the carbon atoms. While for the Sun-CNTs they present us in their atomistic configuration, a porous nanotube (see Fig. \ref{FIG:Irida-CNT:01} and \ref{FIG:Sun-CNT:02}), respectively. Thus, when subjected to elastic deformation, Irida-CNTs exhibit nanostructural reconstruction between carbon atoms (see Fig. \ref{FIG:Irida-CNT(n,0)(n,n);snapshot_reconstruction_nanostructural:09}) before complete nanofracture. Therefore, the tensile strain of the Irida-CNTs is greater than that of the Sun-CNTs.

\section{Conclusions}

We investigated the nanostructural and elastic properties of Irida-graphene-based nanotubes (Irida-CNTs) and Sun-graphene-based nanotubes (Sun-CNTs) of different radii, chiralities and lengths, through extensive fully atomistic reactive (ReaxFF) classical molecular dynamics simulations. We compared the Young's Modulus values reported in the literature for conventional carbon nanotubes (CNTs). Our results show that the complete nanostructural failure (mechanical fracture) of Irida-CNT has a slightly higher value than Sun-CNTs and occurred around critical strain values ($\epsilon_{c} = 28.10$ \%). This result indicates that Irida-CNTs behave as flexible, high-strength nanostructures due to nanostructural reconstructions, wherein in the stretching regimen some bonds are broken (chemical bonds aligned in the direction of uniaxial stretching in the $z$-direction). With relation to the nanofracture patterns, under stretch, the stress accumulation occurs along lines composed of carbon atoms bonded by covalent bonds parallel to the externally applied strain direction, the region where nanostructural mechanical failure begins up to nanofractured complete of Irida-CNTs and Sun-CNTs. The stress-strain curves for Irida-CNTs and Sun-CNTs with distinct diameters, chirality and length are characterized by the existence of linear (elastic) and plastic regimes, where the bonds start to break and propagate until reaching a complete mechanical nanofracture. The Irida-CNTs are similar to the ones observed in brittle materials and/or nanostructures under load strain while Sun-CNTs are similar to ductile materials and/or nanostructures. The values of Young’s modulus for Irida-CNTs are slightly larger range ($640.34$ GPa - $825.00$ GPa) and Sun-CNTs, the Young’s Modulus range ($200.44$ GPa - $472.84$ GPa). Lower values than those presented for conventional carbon nanotubes (CNT). The difference is explicable in nanostructural morphology as indicating that Sun-CNT is porosity, it is still possible to have carbon nanotubes-based Irida-CNT and Sun-CNT with relatively considerable Young's modulus values.

\section{Acknowledgements}

This work was partly supported by the Brazilian agencies CAPES, CNPq, FAPESP, FAPEPI. J.M.S acknowledges CENAPAD-SP (Centro Nacional de Alto Desempenho em São Paulo - Universidade Estadual de Campinas - UNICAMP) for the computational support process (proj842). The authors R. Kalami, S.A. Ketabi and J.M. De Sousa would like to thank the School of Physics, Damghan University, Condensed Matter Physics department. R.S.F. acknowledges the Laboratório de Simulação Computacional Cajuína (LSCC) at Universidade Federal do Piauí, CENAPAD-CE and CENAPAD-SP, for the computational support. R.S.F. acknowledges the funding from CNPq (Grant no. 406447/2022-5 - Materials Informatics)



\newpage

\bibliographystyle{elsarticle-num}
\bibliography{bibliografia.bib}

\begin{thebibliography}{10}
\expandafter\ifx\csname url\endcsname\relax
  \def\url#1{\texttt{#1}}\fi
\expandafter\ifx\csname urlprefix\endcsname\relax\def\urlprefix{URL }\fi
\expandafter\ifx\csname href\endcsname\relax
  \def\href#1#2{#2} \def\path#1{#1}\fi

\bibitem{iijima1991helical}
S.~Iijima, Helical microtubules of graphitic carbon, nature 354~(6348) (1991)
  56--58.

\bibitem{jin2022cagdal3o7}
M.~Jin, N.~Li, H.~Shao, D.~Li, F.~Sun, W.~Yu, X.~Dong, Cagdal3o7: Eu3+
  unidimensional nanostructures: Facile electrospinning synthesis, structure
  and luminescence, Ceramics International 48~(21) (2022) 31548--31558.

\bibitem{campiglio2011quasi}
P.~Campiglio, V.~Repain, C.~Chacon, O.~Fruchart, J.~Lagoute, Y.~Girard,
  S.~Rousset, Quasi unidimensional growth of co nanostructures on a strained au
  (111) surface, Surface science 605~(13-14) (2011) 1165--1169.

\bibitem{herrera2021theoretical}
A.~Herrera-Carbajal, V.~Rodr{\'\i}guez-Lugo, J.~Hern{\'a}ndez-{\'A}vila,
  A.~S{\'a}nchez-Castillo, A theoretical study on the electronic, structural
  and optical properties of armchair, zigzag and chiral silicon--germanium
  nanotubes, Physical Chemistry Chemical Physics 23~(23) (2021) 13075--13086.

\bibitem{dresselhaus1998physical}
G.~Dresselhaus, M.~S. Dresselhaus, R.~Saito, Physical properties of carbon
  nanotubes, World scientific, 1998.

\bibitem{reich2008carbon}
S.~Reich, C.~Thomsen, J.~Maultzsch, Carbon nanotubes: basic concepts and
  physical properties, John Wiley \& Sons, 2008.

\bibitem{ebbesen1994carbon}
T.~W. Ebbesen, Carbon nanotubes, Annual review of materials science 24~(1)
  (1994) 235--264.

\bibitem{salvetat2002mechanical}
J.-P. Salvetat-Delmotte, A.~Rubio, Mechanical properties of carbon nanotubes: a
  fiber digest for beginners, Carbon 40~(10) (2002) 1729--1734.

\bibitem{bhattacharyya2008carbon}
S.~Bhattacharyya, S.~Guillot, H.~Dabboue, J.-F. Tranchant, J.-P. Salvetat,
  Carbon nanotubes as structural nanofibers for hyaluronic acid hydrogel
  scaffolds, Biomacromolecules 9~(2) (2008) 505--509.

\bibitem{zhang2010functional}
W.-D. Zhang, B.~Xu, L.-C. Jiang, Functional hybrid materials based on carbon
  nanotubes and metal oxides, Journal of Materials Chemistry 20~(31) (2010)
  6383--6391.

\bibitem{navrotskaya2020hybrid}
A.~G. Navrotskaya, D.~D. Aleksandrova, E.~F. Krivoshapkina,
  M.~Sillanp{\"a}{\"a}, P.~V. Krivoshapkin, Hybrid materials based on carbon
  nanotubes and nanofibers for environmental applications, Frontiers in
  Chemistry 8 (2020) 546.

\bibitem{legoas2003molecular}
S.~Legoas, V.~Coluci, S.~Braga, P.~Coura, S.~Dantas, D.~S. Galvao,
  Molecular-dynamics simulations of carbon nanotubes as gigahertz oscillators,
  Physical review letters 90~(5) (2003) 055504.

\bibitem{coelho2015carbon}
C.~Coelho, A.~T. Sep{\'u}lveda, L.~A.~M. Rocha, A.~Ferreira~da Silva, Carbon
  nanotubes: the challenges of the first syntheses trials (2015).

\bibitem{sharma2016biomedical}
P.~Sharma, N.~Kumar~Mehra, K.~Jain, N.~Jain, Biomedical applications of carbon
  nanotubes: a critical review, Current drug delivery 13~(6) (2016) 796--817.

\bibitem{saliev2019advances}
T.~Saliev, The advances in biomedical applications of carbon nanotubes, C 5~(2)
  (2019) 29.

\bibitem{de1995carbon}
W.~A. De~Heer, A.~Chatelain, D.~Ugarte, A carbon nanotube field-emission
  electron source, science 270~(5239) (1995) 1179--1180.

\bibitem{harrison2007carbon}
B.~S. Harrison, A.~Atala, Carbon nanotube applications for tissue engineering,
  Biomaterials 28~(2) (2007) 344--353.

\bibitem{baughman1999carbon}
R.~H. Baughman, C.~Cui, A.~A. Zakhidov, Z.~Iqbal, J.~N. Barisci, G.~M. Spinks,
  G.~G. Wallace, A.~Mazzoldi, D.~De~Rossi, A.~G. Rinzler, et~al., Carbon
  nanotube actuators, Science 284~(5418) (1999) 1340--1344.

\bibitem{lima2012electrically}
M.~D. Lima, N.~Li, M.~Jung~de Andrade, S.~Fang, J.~Oh, G.~M. Spinks, M.~E.
  Kozlov, C.~S. Haines, D.~Suh, J.~Foroughi, et~al., Electrically, chemically,
  and photonically powered torsional and tensile actuation of hybrid carbon
  nanotube yarn muscles, science 338~(6109) (2012) 928--932.

\bibitem{ouyang2002fundamental}
M.~Ouyang, J.-L. Huang, C.~M. Lieber, Fundamental electronic properties and
  applications of single-walled carbon nanotubes, Accounts of chemical research
  35~(12) (2002) 1018--1025.

\bibitem{lu1997elastic}
J.~P. Lu, Elastic properties of carbon nanotubes and nanoropes, Physical review
  letters 79~(7) (1997) 1297.

\bibitem{baughman1987structure}
R.~Baughman, H.~Eckhardt, M.~Kertesz, Structure-property predictions for new
  planar forms of carbon: Layered phases containing sp 2 and sp atoms, The
  Journal of chemical physics 87~(11) (1987) 6687--6699.

\bibitem{coluci2003families}
V.~Coluci, S.~Braga, S.~Legoas, D.~Galvao, R.~Baughman, Families of carbon
  nanotubes: Graphyne-based nanotubes, Physical Review B 68~(3) (2003) 035430.

\bibitem{coluci2004theoretical}
V.~Coluci, D.~Galvao, R.~Baughman, Theoretical investigation of
  electromechanical effects for graphyne carbon nanotubes, The Journal of
  chemical physics 121~(7) (2004) 3228--3237.

\bibitem{de2019elastic}
J.~M. De~Sousa, R.~Bizao, V.~Sousa~Filho, A.~Aguiar, V.~Coluci, N.~Pugno,
  E.~Girao, A.~Souza~Filho, D.~Galvao, Elastic properties of graphyne-based
  nanotubes, Computational materials science 170 (2019) 109153.

\bibitem{de2016torsional}
J.~M. de~Sousa, G.~Brunetto, V.~R. Coluci, D.~S. Galvao, Torsional
  “superplasticity” of graphyne nanotubes, Carbon 96 (2016) 14--19.

\bibitem{zhang2015penta}
S.~Zhang, J.~Zhou, Q.~Wang, X.~Chen, Y.~Kawazoe, P.~Jena, Penta-graphene: A new
  carbon allotrope, Proceedings of the National Academy of Sciences 112~(8)
  (2015) 2372--2377.

\bibitem{de2021mechanical}
J.~M. De~Sousa, A.~Aguiar, E.~Girao, A.~F. Fonseca, V.~Coluci, D.~Galvao,
  Mechanical properties of single-walled penta-graphene-based nanotubes: a dft
  and classical molecular dynamics study, Chemical Physics 547 (2021) 111187.

\bibitem{wang2018popgraphene}
S.~Wang, B.~Yang, H.~Chen, E.~Ruckenstein, Popgraphene: a new 2d planar carbon
  allotrope composed of 5--8--5 carbon rings for high-performance lithium-ion
  battery anodes from bottom-up programming, Journal of Materials Chemistry A
  6~(16) (2018) 6815--6821.

\bibitem{brandao2021mechanical}
W.~Brand{\~a}o, A.~Aguiar, L.~Ribeiro, D.~Galv{\~a}o, J.~M. De~Sousa, On the
  mechanical properties of popgraphene-based nanotubes: a reactive molecular
  dynamics study, ChemPhysChem 22~(7) (2021) 701--707.

\bibitem{wang2015phagraphene}
Z.~Wang, X.-F. Zhou, X.~Zhang, Q.~Zhu, H.~Dong, M.~Zhao, A.~R. Oganov,
  Phagraphene: a low-energy graphene allotrope composed of 5--6--7 carbon rings
  with distorted dirac cones, Nano letters 15~(9) (2015) 6182--6186.

\bibitem{junior2020elastic}
M.~P. J{\'u}nior, J.~M. De~Sousa, W.~Brandao, A.~Aguiar, R.~Bizao, L.~R.
  J{\'u}nior, D.~Galvao, On the elastic properties of single-walled phagraphene
  nanotubes, Chemical Physics Letters 756 (2020) 137830.

\bibitem{novoselov2004electric}
K.~S. Novoselov, A.~K. Geim, S.~V. Morozov, D.-e. Jiang, Y.~Zhang, S.~V.
  Dubonos, I.~V. Grigorieva, A.~A. Firsov, Electric field effect in atomically
  thin carbon films, science 306~(5696) (2004) 666--669.

\bibitem{withers2010electron}
F.~Withers, M.~Dubois, A.~K. Savchenko, Electron properties of fluorinated
  single-layer graphene transistors, Physical review B 82~(7) (2010) 073403.

\bibitem{junior2023irida}
M.~P. J{\'u}nior, W.~da~Cunha, W.~Giozza, R.~de~Sousa~Junior, L.~R. Junior,
  Irida-graphene: A new 2d carbon allotrope, FlatChem (2023) 100469.

\bibitem{tromer2023sun}
R.~M. Tromer, M.~L.~P. Junior, K.~A. Lima, A.~F. Fonseca, D.~S. Galvao,
  L.~A.~R. Junior, Sun-graphyne: A new 2d carbon allotrope with dirac cones,
  arXiv preprint arXiv:2302.08364 (2023).

\bibitem{plimpton1995fast}
S.~Plimpton, Fast parallel algorithms for short-range molecular dynamics,
  Journal of computational physics 117~(1) (1995) 1--19.

\bibitem{van2001reaxff}
A.~C. Van~Duin, S.~Dasgupta, F.~Lorant, W.~A. Goddard, Reaxff: a reactive force
  field for hydrocarbons, The Journal of Physical Chemistry A 105~(41) (2001)
  9396--9409.

\bibitem{mueller2010development}
J.~E. Mueller, A.~C. Van~Duin, W.~A. Goddard~III, Development and validation of
  reaxff reactive force field for hydrocarbon chemistry catalyzed by nickel,
  The Journal of Physical Chemistry C 114~(11) (2010) 4939--4949.

\bibitem{van2003reaxffsio}
A.~C. Van~Duin, A.~Strachan, S.~Stewman, Q.~Zhang, X.~Xu, W.~A. Goddard,
  Reaxffsio reactive force field for silicon and silicon oxide systems, The
  Journal of Physical Chemistry A 107~(19) (2003) 3803--3811.

\bibitem{strachan2003shock}
A.~Strachan, A.~C. van Duin, D.~Chakraborty, S.~Dasgupta, W.~A. Goddard~III,
  Shock waves in high-energy materials: The initial chemical events in
  nitramine rdx, Physical Review Letters 91~(9) (2003) 098301.

\bibitem{strachan2005thermal}
A.~Strachan, E.~M. Kober, A.~C. Van~Duin, J.~Oxgaard, W.~A. Goddard, Thermal
  decomposition of rdx from reactive molecular dynamics, The Journal of
  chemical physics 122~(5) (2005).

\bibitem{senftle2016reaxff}
T.~P. Senftle, S.~Hong, M.~M. Islam, S.~B. Kylasa, Y.~Zheng, Y.~K. Shin,
  C.~Junkermeier, R.~Engel-Herbert, M.~J. Janik, H.~M. Aktulga, et~al., The
  reaxff reactive force-field: development, applications and future directions,
  npj Computational Materials 2~(1) (2016) 1--14.

\bibitem{humphrey1996vmd}
W.~Humphrey, A.~Dalke, K.~Schulten, Vmd: visual molecular dynamics, Journal of
  molecular graphics 14~(1) (1996) 33--38.

\bibitem{martys1999velocity}
N.~S. Martys, R.~D. Mountain, Velocity verlet algorithm for
  dissipative-particle-dynamics-based models of suspensions, Physical Review E
  59~(3) (1999) 3733.

\bibitem{evans1983isothermal}
D.~J. Evans, G.~P. Morriss, The isothermal/isobaric molecular dynamics
  ensemble, Physics Letters A 98~(8-9) (1983) 433--436.

\bibitem{hoover1985canonical}
W.~G. Hoover, Canonical dynamics: Equilibrium phase-space distributions,
  Physical review A 31~(3) (1985) 1695.

\bibitem{mcquarrie1987virial}
D.~McQuarrie, J.~Rowlinson, The virial expansion of the grand potential at
  spherical and planar walls, Molecular Physics 60~(5) (1987) 977--989.

\bibitem{de2016mechanical}
J.~M. De~Sousa, T.~Botari, E.~Perim, R.~Bizao, D.~S. Galvao, Mechanical and
  structural properties of graphene-like carbon nitride sheets, RSC advances
  6~(80) (2016) 76915--76921.

\bibitem{an2011elucidation}
Q.~An, S.~V. Zybin, W.~A. Goddard~III, A.~Jaramillo-Botero, M.~Blanco, S.-N.
  Luo, Elucidation of the dynamics for hot-spot initiation at nonuniform
  interfaces of highly shocked materials, Physical Review B 84~(22) (2011)
  220101.

\bibitem{lu2007inverse}
J.~Lu, X.~Zhou, M.~L. Raghavan, Inverse elastostatic stress analysis in
  pre-deformed biological structures: demonstration using abdominal aortic
  aneurysms, Journal of biomechanics 40~(3) (2007) 693--696.

\bibitem{zienkiewicz1969elasto}
O.~Zienkiewicz, S.~Valliappan, I.~King, Elasto-plastic solutions of engineering
  problems ‘initial stress’, finite element approach, International Journal
  for Numerical Methods in Engineering 1~(1) (1969) 75--100.

\bibitem{mian2005laser}
A.~Mian, G.~Newaz, L.~Vendra, N.~Rahman, D.~Georgiev, G.~Auner, R.~Witte,
  H.~Herfurth, Laser bonded microjoints between titanium and polyimide for
  applications in medical implants, Journal of Materials Science: Materials in
  Medicine 16 (2005) 229--237.

\bibitem{qin2019optimization}
M.~Qin, Z.~Zhang, Y.~Zhao, L.~Liu, B.~Jia, K.~Han, H.~Wu, Y.~Liu, L.~Wang,
  X.~Min, et~al., Optimization of von mises stress distribution in mesoporous
  $\alpha$-fe2o3/c hollow bowls synergistically boosts gravimetric/volumetric
  capacity and high-rate stability in alkali-ion batteries, Advanced Functional
  Materials 29~(34) (2019) 1902822.

\bibitem{goel2014molecular}
S.~Goel, S.~S. Joshi, G.~Abdelal, A.~Agrawal, Molecular dynamics simulation of
  nanoindentation of fe3c and fe4c, Materials Science and Engineering: A 597
  (2014) 331--341.

\bibitem{fago2004density}
M.~Fago, R.~L. Hayes, E.~A. Carter, M.~Ortiz, Density-functional-theory-based
  local quasicontinuum method: Prediction of dislocation nucleation, Physical
  Review B 70~(10) (2004) 100102.

\bibitem{wang2005size}
L.~Wang, Q.~Zheng, J.~Z. Liu, Q.~Jiang, Size dependence of the thin-shell model
  for carbon nanotubes, Physical review letters 95~(10) (2005) 105501.

\bibitem{brandao2023first}
W.~Brand{\~a}o, J.~M. De~Sousa, A.~Aguiar, D.~Galv{\~a}o, L.~A. Ribeiro~Jr,
  A.~F. Fonseca, First-principles and reactive molecular dynamics study of the
  elastic properties of pentahexoctite-based nanotubes, Mechanics of Materials
  183 (2023) 104694.

\bibitem{krishnan1998young}
A.~Krishnan, E.~Dujardin, T.~Ebbesen, P.~Yianilos, M.~Treacy, Young’s modulus
  of single-walled nanotubes, Physical review B 58~(20) (1998) 14013.

\bibitem{michal2012mechanical}
B.~Micha{\l}, R.~Jaros{\l}aw, Mechanical properties of single-walled carbon
  nanotubes simulated with airebo force-field, CMST 18~(2) (2012) 67--77.

\end{thebibliography}

\newpage

\begin{table}[hbt!]
    \centering
    \caption{We present the nanostructural parameters of the Irida-G-NTs and Sun-G-NTs studied here by reactive classical molecular dynamics simulations method.}
    \begin{tabular}{|c|c|c|c|c|c|}
    \hline
    \multicolumn{6}{|c|}{Irida-graphene-based nanotubes (Irida-CNTs)} \\ \hline
     chirality & Irida-CNTs & number of atoms & radius (\AA) & diameter (\AA) & length (\AA)  \\ \hline
      \multirow{5}{*}{$(n,0)$}
        & $(5,0)$   & 600 & 5.045  & 10.089 & 54.900 \\ \cline{2-6}
        & $(6,0)$  & 720 &6.054  & 12.107 & 54.900 \\ \cline{2-6}
        & $(7,0)$  & 840 & 7.063  & 14.125 & 54.900 \\ \cline{2-6}
        & $(8,0)$  & 960 &8.071 & 16.143 & 54.900 \\ \cline{2-6}
        & $(9,0)$ & 1080 & 9.080 & 18.161 & 54.900 \\  \cline{2-6}
        & $(10,0)$ & 1200 & 10.089 & 20.179 & 54.900 \\ \hline \cline{2-6}
       \multirow{5}{*}{$(n,n)$} 
        & $(3,3)$   & 576 &  5.243 & 10.485 & 50.714 \\ \cline{2-6}
        & $(4,4)$  & 768 & 6.990 & 13.980 & 50.714 \\ \cline{2-6}
        & $(5,5)$  & 960 & 8.738 & 17.475 & 50.714 \\ \cline{2-6}
        & $(6,6)$  & 1152 & 10.485 & 20.970 & 50.714 \\ \cline{2-6}
        & $(7,7)$  & 1344 & 12.233 & 24.465 & 50.714 \\ \cline{2-6}
        & $(8,8)$ & 1536 & 13.980 &27.960&50.714 \\ \hline
        \multicolumn{6}{|c|}{Sun-graphene-based nanotubes (Sun-CNTs)} \\ \hline
        chirality & Sun-CNTs & number of atoms & radius (\AA) & diameter (\AA) & length (\AA)  \\ \hline
        \multirow{6}{*}{$(n,0)$}
        & $(4,0)$   & 448 & 4.718 & 9.435 & 51.873   \\ \cline{2-6}
        & $(5,0)$   & 560 & 5.897 & 11.794  & 51.873   \\ \cline{2-6}
        & $(6,0)$   & 672 & 7.076 & 14.153 & 51.873  \\ \cline{2-6}
        & $(7,0)$   & 784 & 8.256 & 16.512 & 51.873   \\ \cline{2-6}
        & $(8,0)$   & 896 & 9.435 & 18.871 & 51.873   \\ \cline{2-6}
        & $(9,0)$   &1008 &10.615 & 21.229 & 51.873  \\ \hline
        \multirow{6}{*}{$(n,n)$}
        & $(3,3)$  & 480 & 5.004 & 10.008 & 52.400   \\ \cline{2-6}
        & $(4,4)$  & 640 & 6.672 & 13.344 & 52.400   \\ \cline{2-6}
        & $(5,5)$  & 800 & 8.340 & 16.679 & 52.400   \\ \cline{2-6}
        & $(6,6)$ & 960 &10.008 & 20.015 & 52.400    \\ \cline{2-6} 
        & $(7,7)$ & 1120 & 11.676 & 23.351 & 52.400   \\ \cline{2-6}
        & $(8,8)$ & 1280 & 13.344 & 26.687 & 52.400    \\ \hline
        \multirow{6}{*}{$(n,m)$}
        & $(3,2)$ & 416 & 4.252 & 8.505 & 53.438   \\ \cline{2-6}
        & $(4,3)$ & 800 & 5.897 & 11.794 & 74.105   \\ \cline{2-6}
        & $(5,4)$ & 656 & 7.552 & 15.104 & 47.450   \\ \cline{2-6}
        & $(6,5)$ & 976 & 9.212 & 18.423 & 57.878   \\ \cline{2-6}
        & $(7,6)$ & 1360 &10.874 & 21.747 & 68.321   \\ \cline{2-6}
        & $(8,7)$ & 1808 & 12.537  & 25.075 & 78.774   \\ \cline{2-6} \hline

    \end{tabular}
        \label{tab:Irida-CNT;Sun-CNT}
\end{table}

\begin{table}[hbt!]
    \centering
        \caption{The Elasticity parameters for Irida-graphene-based (Irida-CNT) and Sun-graphene-based (Sun-CNT) nanotubes at room temperature studied in this work. Values of the Young's Modulus (GPa), Ultimate Tensile Strength - UTS (GPa) and critical strain (\%) obtained by reactive (ReaxFF) Classical Molecular Dynamics Simulations. The linear regime considered was $3$\%~ of load strain.}
    \begin{tabular}{|c|c|c|c|c|c|}
    \hline
    \multicolumn{6}{|c|}{Irida-graphene-based nanotubes (Irida-CNTs)} \\ \hline
     chirality & Irida-CNTs & number of atoms & $Y_{mod}$ (GPa) & UTS (GPa) & $\varepsilon_C$ (\%)  \\ \hline
      \multirow{5}{*}{$(n,0)$}
        & $(5,0)$   & 600 & 640.34 $\pm$ 17.62  & 104.19  & 13.69 \\ \cline{2-6}
        & $(6,0)$  & 720 &655.06  $\pm$ 19.49 & 99.75 & 16.82 \\ \cline{2-6}
        & $(7,0)$  & 840 & 661.85 $\pm$ 7.86  & 100.13 & 17.80 \\ \cline{2-6}
        & $(8,0)$  & 960 &682.48 $\pm$ 9.07& 100.64 & 18.98 \\ \cline{2-6}
        & $(9,0)$ & 1080 & 695.95$\pm$ 6.57 & 98.72 & 19.47 \\  \cline{2-6}
        & $(10,0)$ & 1200 & 708.43 $\pm$ 6.63& 96.75 & 19.18 \\ \hline \cline{2-6}
       \multirow{5}{*}{$(n,n)$} 
        & $(3,3)$   & 576 &  737.14 $\pm$ 12.39& 93.33 & 27.90 \\ \cline{2-6}
        & $(4,4)$  & 768 & 773.38 $\pm$ 11.79& 102.10 & 28.00 \\ \cline{2-6}
        & $(5,5)$  & 960 & 785.30 $\pm$ 9.77& 104.42 & 27.63 \\ \cline{2-6}
        & $(6,6)$  & 1152 & 868.23 $\pm$ 8.16& 107.95 & 28.10 \\ \cline{2-6}
        & $(7,7)$  & 1344 & 828.80 $\pm$ 7.66& 109.11 & 27.90 \\ \cline{2-6}
        & $(8,8)$ & 1536 & 825.00 $\pm$ 6.94 &104.29&27.38 \\ \hline
        \multicolumn{6}{|c|}{Sun-graphene-based nanotubes (Sun-CNTs)} \\ \hline
        chirality & Sun-CNTs & number of atoms & $Y_{mod}$ (GPa) & UTS (GPa) & $\varepsilon_C$ (\%)  \\ \hline
        \multirow{6}{*}{$(n,0)$}
        & $(4,0)$   & 448 & 371.39 $\pm$ 7.38 & 73.68 & 18.49   \\ \cline{2-6}
        & $(5,0)$   & 560 & 424.95 $\pm$ 7.77 & 81.95  & 18.81   \\ \cline{2-6}
        & $(6,0)$   & 672 & 432.67 $\pm$ 8.50 & 86.42 & 19.08  \\ \cline{2-6}
        & $(7,0)$   & 784 & 437.46 $\pm$ 7.47 & 86.85 & 19.67   \\ \cline{2-6}
        & $(8,0)$   & 896 & 439.96 $\pm$ 9.10 & 90.48 & 19.47   \\ \cline{2-6}
        & $(9,0)$   &1008 & 472.84 $\pm$ 5.76 & 104.52 & 24.67  \\ \hline
        \multirow{6}{*}{$(n,n)$}
        & $(3,3)$  & 480 & 201.27 $\pm$ 8.50 & 88.10 & 19.47   \\ \cline{2-6}
        & $(4,4)$  & 640 & 228.12 $\pm$ 6.54 & 84.99 & 19.76   \\ \cline{2-6}
        & $(5,5)$  & 800 & 226.58 $\pm$ 5.54 & 87.28 & 19.86   \\ \cline{2-6}
        & $(6,6)$ & 960 &223.63 $\pm$ 5.38 & 85.25 & 19.96    \\ \cline{2-6} 
        & $(7,7)$ & 1120 & 200.44 $\pm$ 5.55 & 87.39 & 20.00   \\ \cline{2-6}
        & $(8,8)$ & 1280 & 214.20 $\pm$ 4.82 & 86.00 & 20.35    \\ \hline
        \multirow{6}{*}{$(n,m)$}
        & $(3,2)$ & 416 & 229.27 $\pm$ 5.47 & 88.10 & 19.47   \\ \cline{2-6}
        & $(4,3)$ & 800 & 231.31 $\pm$ 5.12 & 81.93 & 21.82   \\ \cline{2-6}
        & $(5,4)$ & 656 & 230.37 $\pm$ 5.88 & 84.67 & 20.84   \\ \cline{2-6}
        & $(6,5)$ & 976 & 218.49 $\pm$ 5.45 & 85.33 & 20.16   \\ \cline{2-6}
        & $(7,6)$ & 1360 &214.65 $\pm$ 4.12 & 81.87 & 21.63   \\ \cline{2-6}
        & $(8,7)$ & 1808 & 203.53 $\pm$ 4.04  & 83.65 & 19.96   \\ \cline{2-6} \hline  
    \end{tabular}
    \label{Table:Values:Results:01:Irida-CNT;Sun-CNT}
\end{table}

\end{document}